\documentclass[pra,showpacs,aps,twocolumn,epsfig]{revtex4}

\usepackage{psfrag,graphicx}
\usepackage{amsmath,amssymb}
\newcommand{\ket}[1]{\left| #1 \right>}
\newcommand{\bra}[1]{\left< #1 \right|}

\begin{document}
\title{Unconditional security of coherent-state-based differential phase shift quantum key distribution protocol with block-wise phase randomization}
\author{Kiyoshi Tamaki $^{1,2}$}
\author{Masato Koashi$^{3}$}
\author{Go Kato$^{4}$}

\affiliation{
$^{1}$NTT Basic Research Laboratories, NTT Corporation,\\
3-1,Morinosato Wakamiya Atsugi-Shi, Kanagawa, 243-0198, Japan\\
$^{2}$National Institute of Information and Communications Technology, \\
4-2-1 Nukui-Kita, Koganei, Tokyo 184-8795, Japan\\ 
$^{3}$ Photon Science Center, The University of Tokyo,\\
2-11-16 Yayoi, Bunkyo-ku, Tokyo 113-8656, Japan\\
$^{4}$NTT Communication Science Laboratories, NTT Corporation\\
3-1,Morinosato Wakamiya Atsugi-Shi, Kanagawa, 243-0198, Japan\\}
\date{\today}

\begin{abstract}
We prove the unconditional security of coherent-state-based differential
 phase shift quantum key distribution protocol (DPSQKD) with block-wise
 phase randomization. Our proof is based on the conversion of DPSQKD to
 an equivalent entanglement-distillation protocol 
where the
 estimated phase error rate determines the amount of the privacy
 amplification. 
The generated final key has a contribution from events where the sender
 emits two or more photons, indicating the robustness of DPSQKD against 
photon-number-splitting attacks. 
\end{abstract}

\maketitle
\section{Introduction}
Since the first unconditional security proof of quantum key distribution
 (QKD) by Mayers \cite{maybers96}, quantum information theory has deepened
our understanding on security that is guaranteed by quantum mechanics.
One of the rigorous and intuitive security proofs \cite{LC, SP, GLLP} is
 based on the conversion of a given QKD protocol to a mathematically
 equivalent entanglement distillation protocol (EDP) \cite{EDP}, where
 the sender (Alice) and the receiver (Bob) distill qubit pairs in a maximally entangled state 
(Bell state). 
In this proof technique, the amount of the privacy amplification is
determined from the phase error rate of shared qubit pairs prior to 
distillation. Thus the estimation of the phase 
error rate is the central problem in this approach.
One of the
difficulties in the estimation lies in the fact that the eavesdropper
(Eve) may not attack on each pulse independently, but she may 
interact a train of pulses coherently with her single quantum system.
Hence we are not allowed to assume 
 that the state of the pairs shared by Alice and Bob are
 identically and independently distributed (i.i.d.).
Various techniques have been proposed for solving this problem
to treat the pairs as if they are almost i.i.d., 
such as 
the use of the random sampling theorem \cite{LC}, its generalized
 version to quantum mechanics \cite{TKI04}, Azuma's
 inequality \cite{Azuma, Azuma2}, and quantum De Finetti theorem \cite{Renner}.
With these techniques, we are able to focus on the statistics of 
a single pair, 
which
greatly simplifies the proof. The security of many QKD protocols have
 been proven along this line \cite{DLM06}.

There are QKD protocols, however, for which it seems to be difficult to
prove the security with these techniques. Examples include differential
phase shift QKD protocol (DPSQKD) \cite{DPS} and coherent one-way
protocol (COW protocol) \cite{COW}, which are in the family of the
so-called phase distributed protocols. In these protocols, we employ a
train of pulses in coherent states and 
use the relative phases between the adjacent
pulses to encode the bit information of the key. 
The receiver can read them out simply by optically superposing 
two adjacent pulses by using a optical delay line.
Despite this notable simplicity in the implementation, 
it has been expected that these protocols are robust against 
 photon-number splitting (PNS) attacks that have been found to be
 threats against many protocols using weak coherent laser pulses.
On the other hand, the structure of 
encoding on every pair of adjacent pulses is rather a nuisance if one 
tries to prove the security of these protocols.
Since the whole train of pulses are linked together by 
the phase relations, 
we are no longer allowed to work on each pulse separately, and we may have to
work on a larger Hilbert space for the estimation of the phase error
rate, which makes the security proof difficult \cite{COW2}.
So far, the security of DPSQKD with a single-photon source 
was proved in \cite{single-DPS}, but the robustness against 
PNS attacks cannot be deduced from the proof in which there are no
events of multiple photon emission by the sender. 

In this paper, with no assumption on the attacks available to an 
eavesdropper, we prove the security of DPSQKD with trains
of weak coherent pulses employing block-wise phase
randomization, where Alice applies a common random phase shift 
on every pulse in a block of fixed number $n$ of pulses.
This allows us to use Azuma's inequality to cope with coherent attacks.
As for Bob, we assume that he uses 
detectors that discriminate among the three cases of the vacuum,
a single photon, and two or more photons.
We calculate the final key rates numerically 
as a function of channel transmission
(distance between Alice and Bob) and the observed bit error rate,
and confirm the expected robustness of the DPSQKD protocol 
against the PNS attacks.
For the security proof, 
we introduce an EDP that is shown to be equivalent to the DPSQKD
protocol, and pursue the relation between the bit error rate,
which can be directly estimated, and the phase error rate, 
which determines a sufficient amount of privacy amplification 
to make the final key secure. Although the relevant Hilbert space 
for communication with $n$ pulses is very large with its dimension 
exponential in $n$, we will show that the symmetry in the protocol 
renders the relevant observables in a block diagonalized form, 
which allows us to work essentially on $n$-by-$n$ matrices.

This paper is organized as follows. In Sec. \ref{protocol},
we describe the DPSQKD protocol in detail and clarify 
the assumptions on the devices used by Alice and Bob.
Then we introduce an alternative protocol (EDP-DPS) 
based on entanglement distillation and show that it is equivalent 
to the DPSQKD protocol. In Sec.~\ref{proof}, we develop 
our security proof and explain how to calculate lower bounds on the
secure key rate. Sec.~\ref{examples} shows explicitly how 
the bit error and the phase error are related to each other, 
depending on the number of photons emitted from the sender.
In Sec.~\ref{keyrates}, we show numerical results on the 
key rates for various values of the channel transmission, 
the bit error rate, and the block length $n$.
Then we summarize this
paper in Sec.~\ref{summary}.

\section{DPSQKD and its equivalence to EDP-DPS}\label{protocol}

In this section, we first describe the protocol of DPSQKD together with our 
assumptions on the photon detectors.
We then argue that the DPSQKD protocol is equivalent to 
a protocol involving entanglement distillation, which we call 
the EDP-DPS protocol. Throughout this paper, we use the following
notations:
\begin{eqnarray}
 \hat{P}(\ket{\phi})\equiv \ket{\phi}\bra{\phi}
\end{eqnarray}
for vector $\ket{\phi}$ that is not necessarily unnormalized, 
and 
\begin{eqnarray}
 \hat{H}\equiv \frac{1}{\sqrt{2}}\sum_{s,s'=0,1} (-1)^{ss'}\ket{s}\bra{s'}
\end{eqnarray}
for the Hadamard operator acting on a qubit.

\subsection{Setup of DPSQKD and assumptions}\label{assumption}

\begin{figure}
\begin{center}
 \includegraphics[scale=0.45]{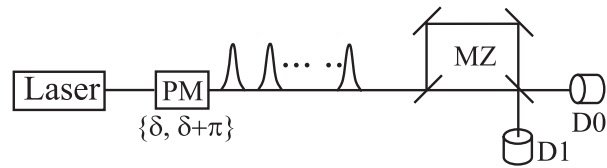}
 \end{center}
 \caption{Schematics of setup for DPSQKD. Alice sends out a train of
 coherent pulses after randomly applying a phase shift,
either $\delta$ or $\delta+\pi$, to each pulse.
Here, $\delta$ is a common offset 
randomly chosen between $0$ and $2\pi$. 
Bob uses a Mach-Zehnder interferometer (depicted as MZ) 
with a delay matched to the interval of the pulses,
followed by photon detectors.
We call the first beam splitter of MZ the shifting beam splitter,
 and the second one the interacting beam splitter. \label{acDPS}}
\end{figure} 

The setup for DPSQKD uses a laser source emitting a long train of pulses 
and a phase shifter at Alice's site, and 
one-bit delay Mach-Zehnder interferometer followed by two photon detectors at Bob's site (see Fig. \ref{acDPS}).
 The interferometer introduces a delay equal to the interval of the
 neighboring pulses, such that photon detection 
at an appropriate timing (we call it $j^{th}$ time slot) 
tells Bob whether 
the relative phase shift between the neighboring pulses 
($j^{th}$ and $(j+1)^{th}$ pulses)
is 0 or $\pi$.
We call the first beam splitter of the interferometer as the shifting beam splitter 
and the second beam splitter as the interacting beam splitter.

Although the protocol described below regards a sequence of $n$ pulses 
as a block, the security analysis does not assume that the pulses
outside of the block should be in the vacuum.
Hence Alice has no need to introduce an optical shutter to 
extract a train composed of exactly $n$ pulses, and 
her source may emit a indefinitely long train of pulses.

Throughout this work, we assume that Bob uses photon-number-resolving
detectors, which can discriminate among the vacuum, a single-photon, and
multi-photons. 
As for the imperfections of the detectors, we assume that 
the inefficiency (a non-unit quantum efficiency) of a detector is modeled by 
a linear absorber followed by a perfect detector with unit quantum efficiency. 
We further assume that 
 the two detectors have the same quantum efficiency.
Then the inefficiency can be modeled by a common 
linear absorber placed in front of the interferometer, which 
may be under Eve's control. As for the dark countings of the detectors,
we assume that they are statistically independent and uniform 
such that it is modeled by Eve's injection of spurious photons
to the interferometer.
Finally, we do not consider any side-channels.

\subsection{DPSQKD}

The protocol of DPSQKD runs as follows.

(1) Alice generates a random $n$-bit sequence 
$\vec{s}\equiv s_1s_2\cdots s_n$ 
and a random (common) phase shift $\delta \in [0,2\pi)$. 
She prepares a sequence of $n$ pulses in the coherent state 
$\bigotimes_{i=1}^n \ket{e^{\mathrm{i}\delta} (-1)^{s_i}\alpha}_i$
and sends them to Bob through a quantum channel. 

(2) Bob receives the $n$ pulses and puts them 
into the shifting beam splitter followed by 
the interacting beam splitter. Then he performs photon
detection by using the photon number resolving detectors. 
Let us call the event {\em detected} if he detects
exactly one photon in $j^{th}(1\le j\le n-1)$ time slot and 
detects the vacuum in all of the other $n$ time slots
(including the $0^{th}$ and $n^{th}$ ones).
If the event is not detected, Alice and Bob skip steps (3) and (4) below.

(3) Bob takes note of the detected bit value and 
announces the number $j$ over the authenticated public channel.

(4) Alice takes note of the bit value $s_j \oplus s_{j+1}$.

(5) Alice and Bob repeat (1) thorough (4) $N$ times. 
Let $QN$ be the number of the detected events.
At this point, Alice and Bob should have their own records of 
$QN$ bits.

(6) 
Alice and  Bob randomly select a small portion $\xi$ of the 
$QN$ detected events, and compare the bit values over 
the public channel. This gives the estimate of the rate $e$
and hence of the number $eQN(1-\xi)$
of bit errors in the remaining portion.

(7) Alice and Bob discuss over the public channel to perform error
correction and privacy amplification on the remaining portion
to share a final key of length $GN(1-\xi)$.

\subsection{Alternative procedures}

\begin{figure}
\begin{center}
 \includegraphics[scale=0.75]{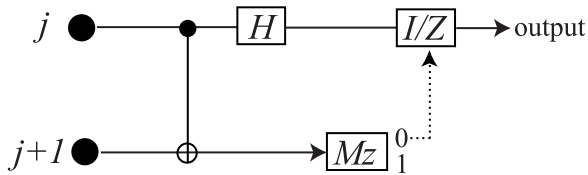}
 \end{center}
 \caption{Schematics of Alice's quantum circuit.
The inputs are the $j^{th}$ and $(j+1)^{th}$ qubits, which
are subjected to a C-NOT gate followed by an Hadamard gate ($H$) on 
the $j^{th}$ qubit. The $(j+1)^{th}$ qubit is then measured on
$Z$-basis ($M_Z$). If the measurement outcome is $0$, a
phase flip gate is applied to the $j^{th}$ qubit with probability 
$1/2$, which is denoted by $I/Z$.
If the outcome is $1$, the $j^{th}$ qubit is left untouched.
\label{QC}}
\end{figure} 

Here we consider equivalent ways of carrying out 
various procedures appearing in the DPSQKD protocol,
in order to reduce it to a protocol based on 
an entanglement distillation
protocol (EDP). 
Suppose that Alice has a quantum register of $n$ qubits.
Let ${\cal H}_A=\bigotimes_{i=1}^n {\cal H}_{A,i}$ be the Hilbert 
space for the $n$ qubits, and define a state 
$$
\ket{\Phi'_\delta}\equiv 2^{-n/2}\sum_{\vec{s}}\bigotimes_{i=1}^n (\hat{H}\ket{s_i}_{A,i})\ket{e^{\mathrm{i}\delta} (-1)^{s_i}\alpha}_i
$$
over the $n$ qubits and the 
$n$ optical pulses.
Alice's procedure at step (1) in the DPSQKD can be equivalently
done by preparing state $\ket{\Phi'_\delta}$ with random $\delta$
and by extracting $\vec{s}$ through a measurement on the $n$ qubits.
Alternatively, let us introduce an auxiliary system $c$ 
with a Hilbert space spanned by an orthonormal basis $\{\nu\}_{\nu=0}^\infty$
and define 
\begin{equation}
\ket{\Phi}\equiv 2^{-n/2}
\sum_{\vec{s}}
\sum_{\nu=0}^\infty \ket{\nu}_c\hat{\pi}_\nu
\bigotimes_{i=1}^n
(\hat{H}\ket{s_i}_{A,i})
\ket{ (-1)^{s_i}\alpha}_i,
\label{eq:koashi-1}
\end{equation}
where $\hat{\pi}_\nu$ is the projection onto the subspace 
for which the total photon number in the $n$ optical pulses is $\nu$.
Alice can carry out step (1) starting with state $\ket{\Phi}$ instead of $\ket{\Phi'_\delta}$
since 
$$
\frac{1}{2\pi}\int_0^{2\pi} d\delta \ket{\Phi'_\delta}\bra{\Phi'_\delta} 
={\rm Tr}_c \ket{\Phi}\bra{\Phi}
$$
holds.

What Alice eventually needs about the sequence $\vec{s}$ in the DPSQKD
is the single bit value $s_j\oplus s_{j+1}$ in step (4). This bit can be
alternatively extracted by applying a quantum circuit shown in Fig. \ref{QC} 
and measuring the output qubit (which is renamed as $A{\rm q}$)
in the standard ($Z$) basis. Notice that
the random phase flip conditioned on the outcome of the measurement on the 
other qubit has no effect on the $Z$-basis ($\{\ket{0}, \ket{1}\}$-basis) measurement. 
The action of the circuit is also described by a set of 
measurement operators 
$\{\hat{M}^{(j)}_k:
{\cal H}_{A,j}\otimes  {\cal H}_{A,j+1}
\to {\cal H}_{A{\rm q}}\}_{k=1,2,3}$ defined as 
\begin{eqnarray}
\hat{M}^{(j)}_1 &=& 
\hat{H}\ket{0}_{A{\rm q}}
({}_{A,j}\bra{0}{}_{A,j+1}\bra{1})
\nonumber \\
&& + \hat{H}\ket{1}_{A{\rm q}}
({}_{A,j}\bra{1}{}_{A,j+1}\bra{0}),
\nonumber \\
\hat{M}^{(j)}_2 &=& \frac{1}{\sqrt{2}}
\ket{0}_{A{\rm q}} ({}_{A,j}\bra{0}{}_{A,j+1}\bra{0}
+{}_{A,j}\bra{1}{}_{A,j+1}\bra{1}),
\nonumber \\
\hat{M}^{(j)}_3 &=& \frac{1}{\sqrt{2}}
\ket{1}_{A{\rm q}} ({}_{A,j}\bra{0}{}_{A,j+1}\bra{0}
-{}_{A,j}\bra{1}{}_{A,j+1}\bra{1}).
\nonumber \\
&&
\end{eqnarray}

\begin{figure}
\begin{center}
 \includegraphics[scale=0.55]{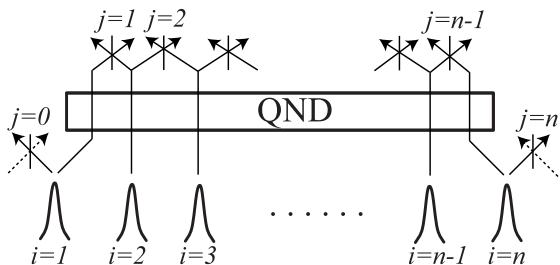}
 \end{center}
 \caption{An equivalent description of Bob's measurement apparatus.
The branching arrows represent the splitting of a pulse by 
the shifting beam splitter of MZ in Fig.~\ref{acDPS}, while 
the intersecting corresponds to the superposition at the 
interacting beam splitter. 
The dotted arrows represent the pulses split from the neighboring blocks. 
The QND measurement in the EDP-DPS is applied to 
the $n$ pulses confined in the box designated as QND in the figure.
Bob's quantum register $B$ is also defined to be the single-photon 
subspace for these $n$ pulses.
\label{DPS}}
\end{figure} 

As for Bob, he can determine whether the event is detected or not,
prior to the determination of the detected time slot and the bit value.
To do this,
suppose that he sends only the $1^{st}$ and the $n^{th}$ pulse to 
the shifting beamsplitters (see Fig. \ref{DPS} ). 
After the $1^{st}$ pulse is split, one of the two halved pulses propagates
toward the interacting beamsplitter of the $1^{st}$ time slot, and 
let us call it the $1^{st}$ half pulse.
Similarly, the $n^{th}$ half pulse goes to the one for 
the $(n-1)^{th}$ time slot. Bob then carries out the 
quantum nondemolition (QND)
measurement 
of the total photon number among the $1^{st}$ half pulse, 
the $n^{th}$ half pulse, and the remaining $(n-2)$ original pulses $(2\le i
\le n-1)$. For the rest of the pulses, he finishes the measurement for
the $0^{th}$ and the $n^{th}$ time slots as in the DPSQKD. 
It is not difficult to see that the event should be {\em detected} 
if and only if the QND measurement reveals exactly one photon
and both the $0^{th}$ and the $n^{th}$ time slots are found in 
the vacuum.

When the event is detected, the state of the $n$ pulses after the 
QND measurement is contained in the subspace ${\cal H}_B$ spanned by 
$n$ states, which we denote  $\{\ket{i}_B\}_{i=1}^n$, 
with $i$ representing the position of the single photon 
(at the half pulse when $i=1,n$ and at the original pulse
otherwise). Let us call this subspace as Bob's quantum register 
$B$. 
Determination of the detected time slot $j$ and the bit value 
in DPSQKD is regarded as a generalized measurement on the 
register $B$. 
Let $\hat{\Pi}_{j,s}$ be the POVM elements for the bit value $s$
detected at the time slot $j$.
Considering the action of the beamsplitters, they are written as
\begin{eqnarray}
\hat{\Pi}_{j,s}=\frac{1}{2}
\hat{P}(\sqrt{\kappa_j}\ket{j}_B+(-1)^s\sqrt{\kappa_{j+1}}\ket{j+1}_B
\label{eq:koashi-def-pijs}
\end{eqnarray}
for $1\le j \le n-1$ with
\begin{eqnarray}
\kappa_1=\kappa_n=1, \;\; \kappa_i=1/2 \;(2\le i \le n-1).
\label{eq:koashi-def-kappa}
\end{eqnarray}

This measurement can be further decomposed into two 
consecutive measurements. The first one is a filtering operation, which gives the outcome $j$
and leaves a qubit ${\cal H}_{B{\rm q}}$. It is described 
by a set of measurement operators $\hat{F}_j:{\cal H}_B \to {\cal H}_{B{\rm
q}}$. The second one measures 
the qubit on the standard basis 
$\{\ket{0}_{B{\rm q}},\ket{1}_{B{\rm q}}\}$.
The entire procedure is equivalent to the POVM $\{\hat{\Pi}_{j,s}\}$
if $\hat{F}_j^\dagger \hat{P}(\ket{s}_{B{\rm q}}) \hat{F}_j=\hat{\Pi}_{j,s}$,
which is satisfied by the following choice
\begin{eqnarray}
\hat{F}_{j} = \sqrt{\kappa_j}\hat{H}\ket{1}_{B{\rm q}} {}_B \bra{j}
+
\sqrt{\kappa_{j+1}}\hat{H}\ket{0}_{B{\rm q}} {}_B \bra{j+1}.
\end{eqnarray}

\subsection{EDP-DPS}

The alternative procedures by Alice and Bob described above lead to 
the following protocol (EDP-DPS) that is equivalent to 
DPSQKD in terms of the security.
 
\mbox{}

(V1) Alice prepares the state $\ket{\Phi}$ and sends the $n$ pulses
to Bob through a quantum channel. 

(V2) Bob receives the $n$ pulses and 
carries out the QND measurement in Fig.~\ref{DPS}
 to test whether there
is exactly one photon in total among the $1^{st}$ half pulse, 
the $n^{th}$ half pulse, and the remaining $(n-2)$ original pulses $(2\le i
\le n-1)$. He also tests whether 
both of the $0^{th}$ and the $n^{th}$ time slot are in the vacuum. 
The event is {\em detected} if both of the tests are qualified.
If the event is not detected, Alice and Bob skip steps (V3) and (V4) below.

(V3)
When the event was detected, Bob's quantum register $B$ is 
measured with measurement operators $\{F_j\}$. He stores the output 
qubit $B{\rm q}$ 
and announces the outcome $j$, which is the position of the detected time
slot, over the authenticated public channel.

(V4) 
Alice applies the quantum circuit in Fig. \ref{QC} to the $j^{th}$
and $(j+1)^{th}$ qubit of her quantum register $A$, 
and stores the output qubit $A{\rm q}$.
She also measures system $c$ to learn the total photon number $\nu$.

(V5) Alice and Bob repeat (V1) through (V4) $N$ times. 
Let $QN$ be the number of the detected events.
At this point, Alice and Bob should share  
$QN$ pairs of qubits.

(V6) 
Alice and Bob randomly select a small portion $\xi$ of the 
$QN$ detected events, measures the qubits in the standard basis, 
and compare the bit values over 
the public channel. This gives the estimate of the rate $e$
and hence of the number $eQN(1-\xi)$
of bit errors in the remaining portion.

(V7) Alice and Bob discuss over the public channel to conduct 
entanglement distillation on the remaining qubit pairs, followed by 
the measurements on the standard basis to 
agree on a final key of length $GN(1-\xi)$.

\mbox{}

Here the equivalence of steps (7) and (V7) follows from the discussion 
in \cite{SP}. The final key rate $G$ is determined through the
statistics of the occurrence of the phase error in the 
$QN(1-\xi)$ pairs in step (V7), which is analyzed in detail 
in the next section.

\section{Security proof of EDP-DPS}\label{proof}

In this section, we prove the security of EDP-DPS. Our proof is based on
the Shor-Preskill security proof where Alice and Bob distill out 
a number of qubit pairs
in a state whose fidelity to as many copies of the Bell state 
$(\ket{0}\ket{0}+\ket{1}\ket{1})/\sqrt{2}$ approaches unity exponentially 
in the number of pairs. The key generated from the distilled state is then secure 
in the sense of composable security \cite{composable}.
According to Shor and Preskill, this virtual distillation protocol can be 
translated to error correction and privacy amplification schemes in the 
actual QKD protocol to obtain a final key of the same security.
The length of the final key is determined from the 
amount of error correction, which reflects the bit error rate,
and from the amount of privacy amplification, which 
is related to the amount of phase errors in the qubit pairs before
the distillation. For simplicity, here we assume the asymptotic limit of large
$N$ in which $\xi$ is negligible, and the efficiency of the error
correction reaches the Shannon limit.
 Then the rate $G$ of the final key 
per transmission of $n$ pulses is written as 
\begin{eqnarray}
G=Q\left[
1-h\left(e
\right)-h^{(\rm ph)}
\right],
\label{eq:koashi-3}
\end{eqnarray}
where $h(x)\equiv -x\log_2 x-(1-x)\log_2(1-x)$.
The number $h^{(\rm ph)}$ implies 
the entropy of the occurrence of 
phase errors, in the sense that 
the number of phase error patterns possible in 
the $l\equiv QN$ qubit pairs prior to distillation 
is given by  
$2^{lh^{(\rm ph)}}$. Our goal is to determine an upper-bound on 
$h^{(\rm ph)}$ from the bit error rate $e$
and the detection rate $Q$.

\subsection{Bit and phase error rates}\label{proof2}

In order to see how the bit errors and the phase errors are related, 
we regard the occurrence of a bit or a phase error as an outcome
of a measurement on Alice and Bob's quantum registers $A$ and $B$
just after the event is detected at step (V2), 
and write down the corresponding POVM elements.
For the occurrence of a bit error from the time slot $j$, the POVM element
is given by
\begin{eqnarray}
 \hat{e}_{j}=\sum_{s,s'}\hat{P}(\hat{H}\ket{s}_{A,j})
\hat{P}(\hat{H}\ket{s'}_{A,j+1})\hat{\Pi}_{j,s\oplus s' \oplus 1}.
\end{eqnarray}
Here and henceforth, we omit identity operators on 
subsystems, like the ones for Alice's $(n-2)$ irrelevant qubits 
in the above expression. 
The occurrence of a phase error is defined 
to be the case where Alice's qubit $A{\rm q}$ and Bob's qubit $B{\rm q}$
produce different outcomes when they are measured on the $X$ basis
$\{\hat{H}\ket{0}, \hat{H}\ket{1}\}$. 
The corresponding POVM element is given by 
\begin{eqnarray}
 \hat{e}_{j}^{{\rm (ph)}}&=&
\sum_{s,k}\hat{M}^{(j)\dagger}_k\hat{P}(\hat{H}\ket{s}_{A{\rm q}})\hat{M}^{(j)}_k
\otimes \hat{F}^{\dagger}_j\hat{P}(\hat{H}\ket{\bar{s}}_{B{\rm q}})\hat{F}_j
\nonumber \\
&=& \sum_{s} \left[\hat{P}(\ket{s}_{A,j}\ket{\bar{s}}_{A,j+1})
+\frac{1}{2}\hat{P}(\ket{0}_{A,j}\ket{0}_{A,j+1})
\right.
\nonumber \\
&& \left. 
+ \frac{1}{2}\hat{P}(\ket{1}_{A,j}\ket{1}_{A,j+1})
\right]
\otimes
\kappa_{j+{s}}\hat{P}(\ket{j+{s}}_B),
\end{eqnarray}
where $\bar{s}\equiv s \oplus 1$.
It is convenient to introduce a unitary operator $\hat{U}$ 
acting on ${\cal H}_A\otimes {\cal H}_B$ defined by 
\begin{eqnarray}
\hat{U}\bigotimes_{i'=1}^n
(\hat{H}\ket{s_{i'}}_{A,i'})
\ket{i}_B
=(-1)^{s_i} 
\bigotimes_{i'=1}^n
(\hat{H}\ket{s_{i'}}_{A,{i'}})
\ket{i}_B
\end{eqnarray}
for $1\le i \le n$.
Then it is straightforward to show that 
\begin{eqnarray}
 \hat{U}\hat{e}_{j}\hat{U}^\dagger=
\hat{\Pi}_{j,1}.
\end{eqnarray}
By using the relation
\begin{eqnarray}
 \hat{U}
\hat{P}(\ket{s}_{A,i})\hat{P}(\ket{i'}_B)\hat{U}^\dagger=
\hat{P}(\ket{s\oplus \delta_{i,i'}}_{A,i})\hat{P}(\ket{i'}_B)
\end{eqnarray}
we also have
\begin{eqnarray}
  \hat{U}\hat{e}_{j}^{{\rm (ph)}}\hat{U}^\dagger &=&
\frac{1}{2}[\hat{P}(\ket{1}_{A,j})
+ \hat{P}(\ket{1}_{A,j+1})]
\nonumber \\
&&
\otimes
[\kappa_{j}\hat{P}(\ket{j}_B)
+\kappa_{j+1}\hat{P}(\ket{j+1}_B)].
\label{eq:koashi-ueph}
\end{eqnarray}

By taking a sum over the time slots, we obtain the operators for 
the bit and phase errors as
\begin{eqnarray}
 \hat{e}\equiv \sum_{j=1}^{n-1} \hat{e}_j, \;\;
 \hat{e}^{{\rm (ph)}}\equiv \sum_{j=1}^{n-1} \hat{e}_j^{{\rm (ph)}}.
\end{eqnarray} 
When the state of Alice and Bob's quantum registers $A$ and $B$
just after the event is detected at step (V2) is $\hat\rho$,
the probability of having a bit error in the extracted qubit pair
$A{\rm q}$ and $B{\rm q}$ is given by ${\rm Tr}(\hat\rho\hat{e})$.
The probability of having a phase error is 
given by ${\rm Tr}(\hat\rho\hat{e}^{{\rm (ph)}})$.
Through the operator $\hat{U}$, these error operators are 
concisely written as follows. For the bit error, we have 
\begin{eqnarray}
 \hat{U}\hat{e}\hat{U}^\dagger&=& \hat{1}_A \otimes 
\hat{\Pi}
\label{eq:koashi-enu}
\\
\hat{\Pi} &\equiv &
\sum_{j=1}^{n-1}\hat{\Pi}_{j,1},
\end{eqnarray}
where $\hat{1}_A$ is the identity operator acting on ${\cal H}_A$.
On the basis $\{\ket{i}_B\}$,
 $\hat{\Pi}$ is represented by a tri-diagonal real symmetric matrix.
From Eq.~(\ref{eq:koashi-def-pijs}), we have 
\begin{eqnarray}
 {}_B\bra{i}\hat\Pi\ket{i}_B&=&1/2\;\;\; (1\le i \le n)
\\
 {}_B\bra{i}\hat\Pi\ket{i+1}_B&=&-1/(2\sqrt{2}) \;\;\; (i=1,n-1)
\\
 {}_B\bra{i}\hat\Pi\ket{i+1}_B&=&-1/4 \;\;\; (2\le i \le n-2).
\end{eqnarray}
Notice that $\hat\Pi\ge 0$, and that 
$\hat\Pi(\sum_i c_i\ket{i}_B)=0$ if and only if 
\begin{eqnarray}
 \sqrt{2}c_1=c_2=\cdots=c_{n-1}=\sqrt{2}c_n.
\label{eq:condition_e_0}
\end{eqnarray}
The factor of $\sqrt{2}$ comes from the fact that quantum register 
$B$ is defined after the $(i=1)$ pulse and the $(i=n)$ pulse
are split in half.

For the phase error,
let us write the standard basis states of Alice's quantum 
register with $n$-bit sequence $\vec{a}\equiv a_1a_2\ldots a_n$
as $\ket{\vec{a}}_A\equiv \ket{a_1}_{A,1}\ket{a_2}_{A,2}\cdots
\ket{a_n}_{A,n}$. 
Then, using Eqs.~(\ref{eq:koashi-def-kappa}) and (\ref{eq:koashi-ueph}), 
we have 
\begin{eqnarray}
 \hat{U}\hat{e}^{{\rm (ph)}}\hat{U}^\dagger=\sum_{\vec{a}}
P(\ket{\vec{a}}_A)\otimes \hat{\Pi}^{(\rm ph)}_{\vec{a}}
\label{eq:koashi-ephnu}
\end{eqnarray}
with 
\begin{eqnarray}
 \hat{\Pi}^{(\rm ph)}_{\vec{a}} &\equiv&
P(\ket{1}_B) \left(
\frac{\delta_{a_{1},1}}{2}+
\frac{\delta_{a_{2},1}}{2}
\right)
\nonumber \\
&+& \sum_{i=2}^{n-1} P(\ket{i}_B)
\left(
\frac{\delta_{a_{i-1},1}}{4}+
\frac{\delta_{a_{i},1}}{2}+
\frac{\delta_{a_{i+1},1}}{4}
\right)
\nonumber \\
&&+P(\ket{n}_B) \left(
\frac{\delta_{a_{n-1},1}}{2}+
\frac{\delta_{a_{n},1}}{2}
\right).
\label{eq:koashi-pipha}
\end{eqnarray}
On the basis $\{\ket{i}_B\}$,
 $\hat{\Pi}$ is tri-diagonal and $\hat{\Pi}^{(\rm ph)}_{\vec{a}}$
is diagonal.

\subsection{Constraints on the state of quantum registers}

Since Alice's quantum register $A$ and 
the system $c$ never leave Alice,
Eve can change their state only indirectly through 
controlling whether the event will be detected or not. 
As a result, if the initial state $\ket{\Phi}$
has zero amplitude for a state vector $\ket{\vec{a}}_A\ket{\nu}_c$
($\bra{\Phi}\ket{\vec{a}}_A\ket{\nu}_c=0$),
it stays so after the event is detected, namely,
its density operator $\hat{\rho}$ satisfies 
$\hat{\rho} \ket{\vec{a}}_A\ket{\nu}_c=0$.
The relation between $\vec{a}$ and $\nu$ for the 
vanishing amplitude is derived as follows.

The initial state $\ket{\Phi}$ in Eq.~(\ref{eq:koashi-1})
is written as 
\begin{eqnarray}
 &&\ket{\Phi}\equiv 2^{-n}
\sum_{\vec{a}}
\ket{\vec{a}}_{A}
\sum_{\nu=0}^\infty
\ket{\nu}_c
\hat{\pi}_\nu
\bigotimes_{i=1}^{n}
(\ket{\alpha}_i+(-1)^{a_i}\ket{-\alpha}_i)
\nonumber \\
&&
\end{eqnarray}
Since the state $\ket{\alpha}-\ket{-\alpha}$ contains 
at least one photon, we have 
\begin{eqnarray}
 {}_A \bra{\vec{a}} {}_c\bra{\nu} \ket{\Phi}=0 \;\; \text{if}
\;\; |\vec{a}|>\nu,
\end{eqnarray}
where $|\vec{a}|$ is the weight of the bit sequence 
$\vec{a}$, namely, the number of 1's in the sequence.
Since the state $\ket{\alpha}+\ket{-\alpha}$ always contains an 
even number of photons and $\ket{\alpha}-\ket{-\alpha}$ contains 
an odd number of photons, we also have
\begin{eqnarray}
 {}_A \bra{\vec{a}} {}_c\bra{\nu} \ket{\Phi}=0 \;\; \text{if}
\;\; (-1)^{|\vec{a}|}\neq (-1)^\nu.
\end{eqnarray}
Therefore, the state of Alice
 and Bob's quantum registers 
after the event was detected 
and system $c$ revealed a photon number $\nu$
is contained in the range of projection operator $\hat{P}^{(\nu)}$
defined as 
\begin{eqnarray}
 \hat{P}^{(\nu)} \equiv  
\sum_{\vec{a}:|\vec{a}|= \nu, \nu-2,\nu-4,\ldots}
\sum_{i=1}^{n}
\hat{P}(\ket{\vec{a}}_A \ket{i}_B).
\end{eqnarray}
The unitary $\hat{U}$ transforms this projection as
\begin{eqnarray}
 \hat{U}\hat{P}^{(\nu)}\hat{U}^\dagger
&=& \sum_{\vec{a}:|\vec{a}|= \nu-1, \nu-3,\ldots}
\hat{P}(\ket{\vec{a}}_A)\otimes \hat{1}_B
\nonumber \\
&&+ \sum_{\vec{a}:|\vec{a}|= \nu+1}
\hat{P}(\ket{\vec{a}}_A) \otimes \hat{P}_{\vec{a}}
\label{eq:koashi-pnu}
\end{eqnarray}
with 
\begin{eqnarray}
 \hat{P}_{\vec{a}}\equiv \sum_{i=1}^{n} \delta_{a_i,1}P(\ket{i}_B).
\end{eqnarray}

\subsection{Relation between the bit and phase errors}

The relation between the bit and phase errors can be 
conveniently expressed through the quantity
$\Omega^{(\nu)}(\lambda)$ for $\lambda\ge 0$, which is defined as 
the largest eigenvalue of the operator
\begin{equation}
\hat{P}^{(\nu)}(\hat{e}^{(\rm ph)}-\lambda \hat{e})\hat{P}^{(\nu)}
\end{equation}
in the range of $\hat{P}^{(\nu)}$. 
The definition ensures that for any state $\hat{\rho}$
in the range of $\hat{P}^{(\nu)}$, the probability of a phase error
is bounded as 
\begin{equation}
{\rm Tr}(\hat{\rho}\hat{e}^{(\rm ph)})
\le \lambda {\rm Tr}(\hat{\rho}\hat{e})+\Omega^{(\nu)}(\lambda)\,.
\label{eq:koashi-10}
\end{equation}

Suppose that at step (V5) of the EDP-DPS, Alice and Bob 
find $NQq^{(\nu)}$ pairs of qubits for which Alice has found 
the total photon number to be $\nu$ at step (V4).
Imagine one sequentially measures these qubit pairs in the Bell basis,
one pair after another, to count the number of bit errors and 
phase errors. Regardless of Eve's attack, the above inequality 
is also true for the conditional 
probabilities of the occurrence of bit and phase errors in the $m$-th
pair, conditioned on the outcomes for the $(m-1)$ preceding pairs, 
since $\hat\rho$ is arbitrary. Then, using Azuma's inequality \cite{Azuma}, 
we are able to show \cite{Azuma2} that a similar inequality holds for 
the total number $NQq^{(\nu)}e^{({\rm ph},\nu)}$ 
of the phase errors and 
the number $NQq^{(\nu)}e^{(\nu)}$ of the bit errors:
\begin{equation}
e^{({\rm ph},\nu)}
\le \lambda e^{(\nu)}+\Omega^{(\nu)}(\lambda)+\epsilon\,.
\nonumber
\end{equation}
When $NQq^{(\nu)}$ gets larger with any fixed value of $\epsilon>0$, 
the probability of this inequality to be violated decreases
exponentially. Here and henceforth, we consider the limit of large 
$N$ and we neglect $\epsilon$ so that we assume
\begin{equation}
e^{({\rm ph},\nu)}
\le \lambda e^{(\nu)}+\Omega^{(\nu)}(\lambda)
\label{eq:koashi-2}
\end{equation}
holds.
 
The above inequality for various values of $\lambda$ determines 
a convex achievable region of $(e^{(\nu)}, e^{({\rm ph},\nu)})$,
which, in principle, determines the 
convex achievable region of $(e^{(\nu)}, h(e^{({\rm ph},\nu)}))$
specified by a set of linear inequalities 
\begin{equation}
h(e^{({\rm ph},\nu)})
\le  \gamma e^{(\nu)} + \Omega^{(\nu)}_h(\gamma)
\label{eq:koashi-4}
\end{equation}
for various values of $\gamma\ge 0$.
The amount of privacy amplification $h^{\rm (ph)}$
appearing in the key rate formula Eq.~(\ref{eq:koashi-3})
is then bounded as 
\begin{equation}
h^{\rm (ph)}= \sum_{\nu=0}^{\infty}q^{(\nu)} h(e^{({\rm ph},\nu)})
\le \gamma e+
\sum_{\nu=0}^{\infty}q^{(\nu)} \Omega^{(\nu)}_h(\gamma)
\label{eq:koashi-5}
\end{equation}
for fixed values of $\{q^{(\nu)}\}$. Here we have used the relation 
$e=\sum_\nu q^{(\nu)} e^{(\nu)}$ for the observed error rate $e$.
The values of $\{q^{(\nu)}\}$ are freely chosen by Eve under the obvious 
constraints from the number of total events where Alice has emitted 
$\nu$ photons:
\begin{equation}
NQq^{(\nu)}\le N e^{-n\alpha^2}\frac{(n\alpha^2)^\nu}{\nu!}.
\end{equation}
As long as the chain of inequalities 
\begin{equation}
\Omega^{(0)}(\lambda) \le \Omega^{(1)}(\lambda) \le
 \Omega^{(2)}(\lambda)
\le \cdots
\label{chain}
\end{equation}
holds for all $\lambda\ge 0$, 
Eve loses nothing by using the events 
with a larger value of $\nu$.
Thus the optimal choice of $\{q^{(\nu)}\}$ to maximize 
the right-hand side of Eq.~(\ref{eq:koashi-5})
is given by 
\begin{eqnarray} 
q^{(\nu)*} \equiv \left\{ \begin{array}{ll} 
Q^{-1}p_\nu
&  (\nu\ge \nu_{\rm min}+1) \\ 
1-Q^{-1}(1-\sum_{\nu'=0}^{\nu_{\rm min}}p_{\nu'})
& (\nu=\nu_{\rm min}) \\ 
0 & (\nu\le \nu_{\rm min}-1),
\end{array} \right. 
\label{qu}
\end{eqnarray}
where $\{p_\nu\}$ is the Poissonian distribution with mean $n\alpha^2$,
\begin{eqnarray}
 p_\nu \equiv e^{-n\alpha^2}\frac{(n\alpha^2)^\nu}{\nu!},
\end{eqnarray} 
and $\nu_{\rm min}$ is the integer satisfying
\begin{eqnarray}
1-\sum_{\nu'=0}^{\nu_{\rm min}}p_{\nu'}< Q
\le 1-\sum_{\nu'=0}^{\nu_{\rm min}-1}p_{\nu'}.
\end{eqnarray} 
On the other hand, the parameter $\gamma$ can be freely 
chosen to obtain the tightest bound. As a result, we
formally obtain an upper bound on $h^{\rm (ph)}$ as 
\begin{equation}
h^{\rm (ph)}\le \min_\gamma
\left[
\gamma e+
\sum_{\nu=0}^{\infty}q^{(\nu)*} \Omega^{(\nu)}_h(\gamma)
\right].
\label{eq:koashi-6}
\end{equation}

In practice, the evaluation of $\Omega^{(\nu)}(\lambda)$ is 
involved for a large $\nu$, as shown in the next subsection.
Hence, for the key rates calculated in this paper,
we used a bound not as tight as Eq.~(\ref{eq:koashi-6}), 
essentially calculating $\Omega^{(\nu)}(\lambda)$ up to $\nu=3$.
The technical detail of this bound is explained in Appendix A.

\subsection{Evaluation of $\Omega^{(\nu)}(\lambda)$}

Here we explain how to calculate the quantity $\Omega^{(\nu)}(\lambda)$
  which is vital for determining the key rate.
Since $\Omega^{(\nu)}(\lambda)$ is the largest eigenvalue of 
$\hat{P}^{(\nu)}(\hat{e}^{(\rm ph)}-\lambda \hat{e})\hat{P}^{(\nu)}$
in the range of $\hat{P}^{(\nu)}$,
it is also
the largest eigenvalue of the operator
\begin{eqnarray}
&& \hat{U} \hat{P}^{(\nu)} \hat{U}^\dagger(
\hat{U}\hat{e}^{(\rm ph)}\hat{U}^\dagger-\lambda \hat{U}
 \hat{e}
\hat{U}^\dagger)\hat{U} \hat{P}^{(\nu)} \hat{U}^\dagger
\nonumber \\
&=&
\sum_{\vec{a}:|\vec{a}|= \nu-1, \nu-3,\ldots}
\hat{P}(\ket{\vec{a}}_A)\otimes (\hat{\Pi}^{({\rm
ph})}_{\vec{a}}-\lambda
\hat\Pi)
\nonumber \\
&&+
 \sum_{\vec{a}:|\vec{a}|= \nu+1}
\hat{P}(\ket{\vec{a}}_A)\otimes
\hat{P}_{\vec{a}}
(\hat{\Pi}^{({\rm
ph})}_{\vec{a}}-\lambda
\hat\Pi)\hat{P}_{\vec{a}},
\end{eqnarray}
in the range of $\hat{U} \hat{P}^{(\nu)} \hat{U}^\dagger$,
where Eqs.~(\ref{eq:koashi-enu}), (\ref{eq:koashi-ephnu}),
and (\ref{eq:koashi-pnu}) are used.
Since it is a direct sum over various operators specified by 
$\vec{a}$, $\Omega^{(\nu)}(\lambda)$ is the largest among 
the eigenvalues of these operators 
in the range of $\hat{U} \hat{P}^{(\nu)} \hat{U}^\dagger$.
Here we may neglect the operators with $|\vec{a}|\le \nu-3$, 
since the definition of $\hat{\Pi}^{({\rm ph})}_{\vec{a}}$ 
in Eq.~(\ref{eq:koashi-pipha}) assures that 
$\hat{\Pi}^{({\rm ph})}_{\vec{a}} \ge \hat{\Pi}^{({\rm ph})}_{\vec{a}'}$
if $a_i\ge a'_i$ for all $i$. 
We thus conclude that $\Omega^{(\nu)}(\lambda)$ is the larger
of the two numbers $\Omega^{(\nu)}_-(\lambda)$ and 
$\Omega^{(\nu)}_+(\lambda)$ defined as follows;
$\Omega^{(\nu)}_-(\lambda)$ is the largest of the eigenvalues 
of the operators
\begin{eqnarray}
\{\hat{\Pi}^{({\rm ph})}_{\vec{a}}-\lambda\hat\Pi \;|\; |\vec{a}|=\nu-1\}.
\end{eqnarray}
Taking $\{\ket{i}_B\}$ as the basis, 
one can calculate $\Omega^{(\nu)}_-(\lambda)$ 
by evaluating the largest eigenvalues of various tri-diagonal 
$n\times n$ matrices designated by $n$-bit sequences 
$\vec{a}$ with $|\vec{a}|=\nu- 1$.
$\Omega^{(\nu)}_+(\lambda)$ is the largest of the eigenvalues 
of the operators
\begin{eqnarray}
  \{\hat{P}_{\vec{a}}(\hat{\Pi}^{({\rm ph})}_{\vec{a}}-\lambda
\hat\Pi)\hat{P}_{\vec{a}} \;|\; |\vec{a}|=\nu+1\}
\end{eqnarray}
in the range of $\hat{P}_{\vec{a}}$.
Since $\hat{\Pi}^{({\rm ph})}_{\vec{a}}-\lambda
\hat\Pi$ is tri-diagonal, 
the off-diagonal element 
${}_B\bra{i'}(\hat{\Pi}^{({\rm ph})}_{\vec{a}}-\lambda\hat\Pi)\ket{i}_B$
vanishes when $|i'-i|\ge 2$. Eq.~(\ref{eq:koashi-pipha})
assures that the element 
${}_B\bra{i''}(\hat{\Pi}^{({\rm ph})}_{\vec{a}}-\lambda\hat\Pi)\ket{i'}_B$
does not depend on $a_{i}$ when $|i'-i|\ge 2$ and $|i''-i|\ge 2$.  
We are thus allowed to focus on a subspace spanned by states
$\ket{i}_B$ with $a_i=1$
for consecutive values of index $i$.
That is to say,
$\Omega^{(\nu)}_+(\lambda)$ is given by the largest of the eigenvalues 
of the operators
\begin{eqnarray}
  \{\hat{P}_{\vec{a}'}(\hat{\Pi}^{({\rm ph})}_{\vec{a}'}-\lambda
\hat\Pi)\hat{P}_{\vec{a}'} \;&|&\; \vec{a}'=\vec{b}_{l,l+k},
0\le k\le \nu,
\nonumber \\
&&
 1\le l, l+k \le n  \}
\label{eq:koashi-mushi}
\end{eqnarray}
in the range of $\hat{P}_{\vec{a}'}$, where 
$\vec{a}'=\vec{b}_{l,k+l}$ means that
$a'_i=0$ for $i<l$ or $k+l<i$, and 
$a'_i=1$ for $l\le i \le l+k$.
One can thus calculate $\Omega^{(\nu)}_+(\lambda)$ 
by evaluating the largest eigenvalues of various tri-diagonal 
$(k+1)\times (k+1)$ matrices with $k\le \nu$.

From a sequence $\vec{a}^*$ and an eigenvector $\sum_i c_{i}^{*}\ket{i}_B$
corresponding to the largest eigenvalue $\Omega^{(\nu)}(\lambda)$,
we can also reconstruct an optimal state of 
quantum registers $AB$ that saturates the inequality 
(\ref{eq:koashi-10}). It is given by 
\begin{eqnarray}
 \hat{U}^\dagger(\ket{\vec{a}^*}_A\otimes \sum_i c_{i}^{*}\ket{i}_B)
=\sum_i c_{i}^{*} \ket{\vec{a}^*+\vec{b}_i}_A\otimes \ket{i}_B,
\label{eq:opt-state1}
\end{eqnarray}
where $\vec{b}_i$ is the sequence for which $|\vec{b}^{(i)}|=1$
and the only `1' is at the $i$-th bit, and the superscript $*$ represents 
that $c_i$ is optimal.

\section{Explicit relations between the bit and phase error rates}
\label{examples}

In the last section, we showed that the final key rate can be determined 
from the knowledge of the relation between the bit error rate 
$e^{(\nu)}$ and the phase error
rate $e^{({\rm ph}, \nu)}$ for each value of photon number 
$\nu$ emitted by Alice. In this section, we explicitly calculate the 
allowed region of $(e^{(\nu)},e^{({\rm ph}, \nu)})$ up to $\nu=3$
by evaluating the function $\Omega^{(\nu)}(\lambda)$ analytically or numerically.


\subsection{Zero-photon part}

First we discuss the trivial case of $\nu=0$, when Alice has emitted no
photons. If we follow the prescription of the last section,
$\Omega^{(0)}(\lambda)=\Omega^{(0)}_+(\lambda)$
since $\Omega^{(0)}_-(\lambda)$ has no candidates for $\nu=0$.
In Eq.~(\ref{eq:koashi-mushi}), the choice $\vec{a}=\vec{b}_{l,l}$
results in ${}_B\bra{l}\hat{\Pi}^{({\rm ph})}_{\vec{a}}\ket{l}_B=1/2$
and ${}_B\bra{l}\hat{\Pi}\ket{l}_B=1/2$ regardless of $l$.
Hence we have 
\begin{eqnarray}
 \Omega^{(0)}(\lambda)=(1-\lambda)/2,
\end{eqnarray}
which corresponds to a single point $(e^{(0)},e^{({\rm ph},0)})=(1/2,1/2)$.
Physically, this is trivial since there should be no correlations 
in any basis in the shared qubit pair when Alice emitted no photons.

\subsection{Single-photon part}

In this subsection 
we consider the case $\nu=1$, where 
a single photon was emitted from Alice.
To calculate $\Omega^{(1)}(\lambda)$,
it is suffice to consider the cases
 $|{\vec a}|=0$ or $|{\vec a}|=2$. 

Since $\hat\Pi_{\vec{a}}^{({\rm ph})}=0$ for $|{\vec a}|=0$,
$\Omega_-^{(1)}(\lambda)$ is given by the largest eigenvalue of 
$-\lambda \hat{\Pi}$, which is zero. This corresponds to the point 
$(e^{(1)},e^{({\rm ph},1)})=(0,0)$, implying that  Eve has done nothing.

When $|{\vec a}|=2$ with $a_{i}=a_{i'}=1$, it can be interpreted as 
Alice has emitted a photon in the $i$-th pulse and
Bob has received a photon in the $i'$-th pulse, or vice versa.
In either case, Eve's attack has moved the location of the single 
photon, which typically occurs when Eve tries to measure the relative
phase between the two pulses. 
Mathematically, nontrivial choices in 
Eq.~(\ref{eq:koashi-mushi}) are $\vec{a}'=\vec{b}_{l,l+1}$,
leading to $(2\times 2)$ matrices
\begin{eqnarray}
\frac{1}{4}
 \begin{pmatrix}
   4-2\lambda & \sqrt{2} \lambda
\\
  \sqrt{2} \lambda & 3-2\lambda
 \end{pmatrix},
\;
\frac{1}{4}
 \begin{pmatrix}
   3-2\lambda &  \lambda
\\
   \lambda & 3-2\lambda
 \end{pmatrix},
\end{eqnarray}
for $l=1,n-1$ and $2\le l \le n-2$, respectively.
The former matrix always has the largest of the eigenvalues,
leading to $\Omega_+^{(1)}(\lambda)=(7-4\lambda+\sqrt{1+8\lambda^2})/8$.
Combined with $\Omega_-^{(1)}(\lambda)=0$, we have
\begin{eqnarray}
 \Omega^{(1)}(\lambda)=
\begin{cases}
  0   &  (\lambda \ge 6)
\\
 (7-4\lambda+\sqrt{1+8\lambda^2})/8 & (\lambda< 6).
\end{cases}
\end{eqnarray}
$\Omega^{(1)}(6)=0$ yields an inequality 
\begin{eqnarray}
 e^{({\rm ph},1)}\le 6e^{(1)},
\label{eq:6e}
\end{eqnarray}
which is saturated for $0\le e^{(1)} \le 5/34$ ($0\le e^{({\rm ph},1)} \le
15/17$). Hence we have 
$h(e^{({\rm ph},1)})\le h(6e^{(1)})$ for $0\le e^{(1)} \le 1/12$.
This is rewritten as 
$h(e^{({\rm ph},1)})\le 6h'(6\tilde{e})(e^{(1)}-\tilde{e})+h(6\tilde{e})$
for a constant $\tilde{e}$ with $0\le \tilde{e} \le 1/12$, which gives 
the form of $\Omega^{(1)}_h(\gamma)$ in Eq.~(\ref{eq:koashi-4})
implicitly as
\begin{eqnarray}
 \Omega^{(1)}_h(\gamma)= h(6\tilde{e}) - \gamma \tilde{e}
\;\; \text{for} \;\; 
\gamma= 6h'(6\tilde{e}).
\label{eq:koashi-omegah1gamma}
\end{eqnarray}

\subsection{Two-photon part}\label{sec:two-photon}

\begin{figure}
\begin{center}
 \includegraphics[scale=0.6]{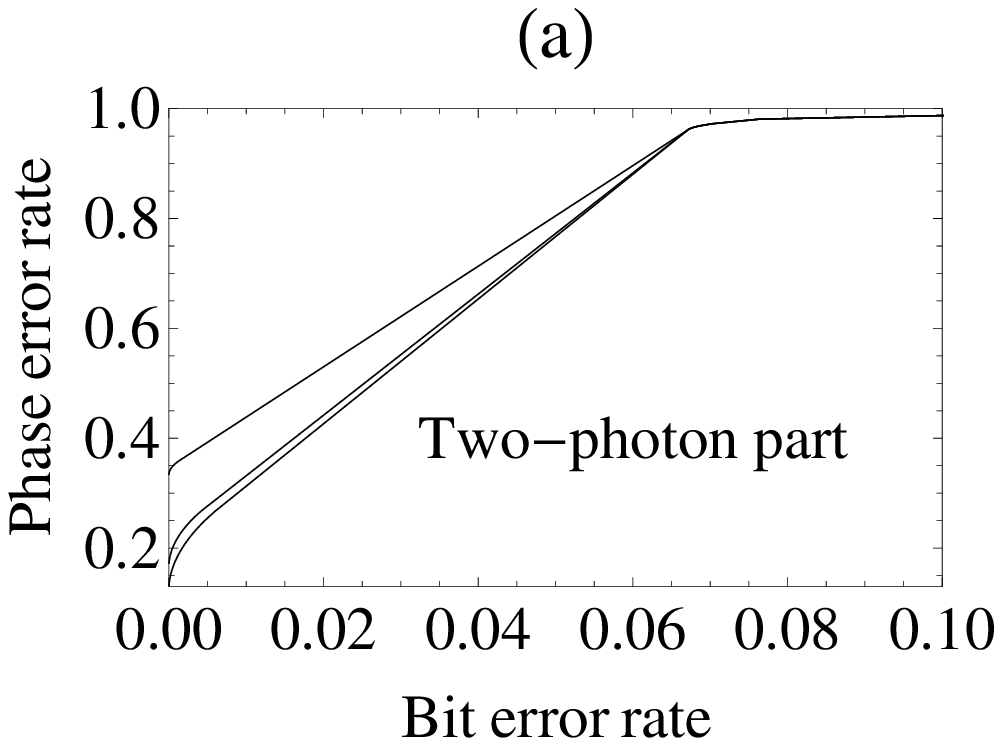}
\includegraphics[scale=0.6]{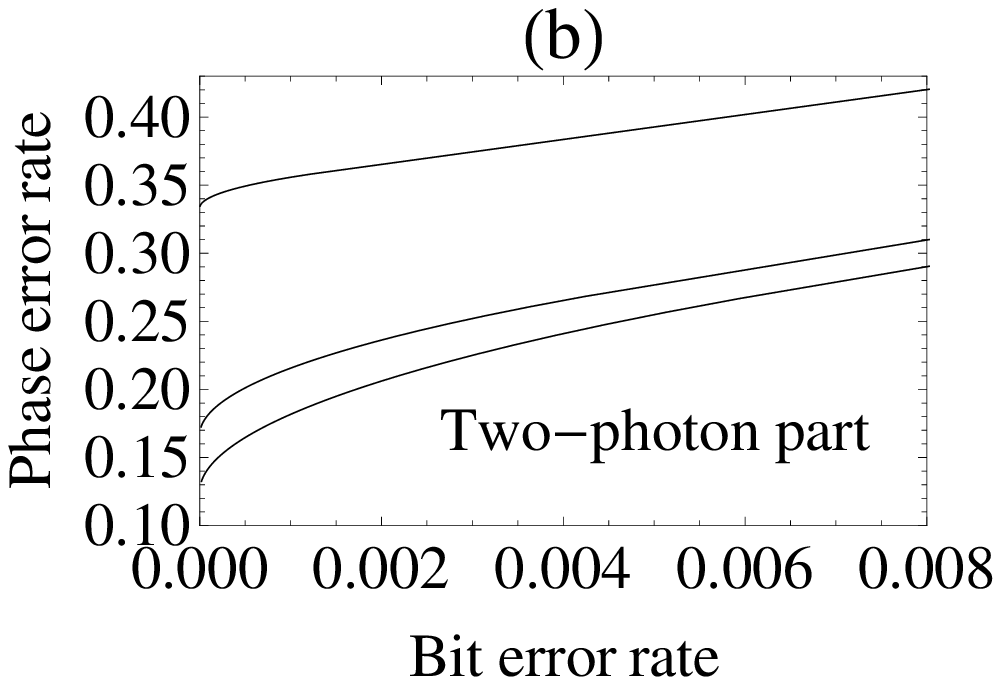}
 \end{center}
 \caption{(a) The upper bound on the phase error rate 
$e^{({\rm ph},2)}$
as a function of the bit error rate $e^{(2)}$
for the two-photon part. From top to bottom, the curves correspond to
 the case of $n=4$, $n=7$, and $n=9$, respectively. 
(b) Magnification of the low bit error region. 
\label{fig:two2}}
\end{figure}

For the cases where Alice emitted two photons ($\nu=2$), 
the evaluation of $\Omega^{(2)}(\lambda)=
\max\{\Omega^{(2)}_+(\lambda), \Omega^{(2)}_-(\lambda)\}$
involves calculation of the largest eigenvalues of 
$n$-by-$n$ matrices, which we have done numerically.
It has turned out that, irrespective of block size $n$, 
$\Omega^{(2)}_+(\lambda)$ is always determined by 
the choice $\vec{a}=11100\ldots 0$ among various values of $\vec{a}$ with 
$|\vec{a}|=3$,
while $\Omega^{(2)}_-(\lambda)$ is determined by 
$\vec{a}=0100\ldots 0$ which gives an eigenvalue no smaller than any other
 $\vec{a}$ with $|\vec{a}|=1$.
 Whether $\Omega^{(2)}_+(\lambda)$ is larger than $\Omega^{(2)}_-(\lambda)$
or not depends on $\lambda$; With a constant $\lambda_0$ which is solely
dependent on $n$, $\Omega^{(2)}_+(\lambda)\le \Omega^{(2)}_-(\lambda)$
for $\lambda\ge \lambda_0$, while 
$\Omega^{(2)}_+(\lambda)> \Omega^{(2)}_-(\lambda)$
for $ \lambda_0> \lambda\ge 0$.
As a result, the boundary of $(e^{(2)},e^{({\rm ph}, 2)})$
consists of two convex curves determined from $\Omega^{(2)}_\pm(\lambda)$
and a straight line with slope $\lambda_0$ connecting them,
which is shown in Fig. \ref{fig:two2} for $n=4, 7, 9$.
Eve's best strategy is thus to use the subspace with 
$\vec{a}=0100\ldots 0$ for smaller values of $e^{(2)}$ and 
$e^{({\rm ph},2)}$ (low-error regime),
while she should mix it with the subspace with 
$\vec{a}=1110\ldots 0$ for achieving
larger values of $e^{(2)}$ and $e^{({\rm ph},2)}$ (high-error regime).
We also see that as $n$ gets larger, the bound on the phase error rate
becomes tighter. 

\begin{figure}
\begin{center}
 \includegraphics[scale=0.6]{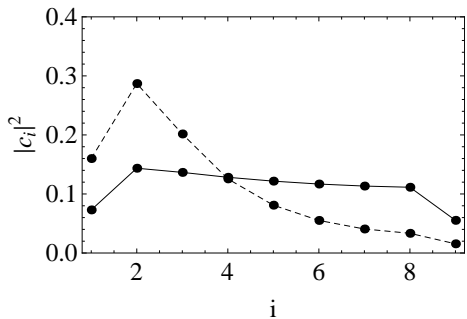}
 \end{center}
 \caption{The optimal values of $\{|c_{i}|^2\}$ for 
$e^{(2)}=0.00010$ (solid lines) and $e^{(2)}=0.00655$ (dashed lines), when $n=9$. \label{fig:two-amp1}}
\end{figure}

\begin{figure}
\begin{center}
 \includegraphics[scale=0.6]{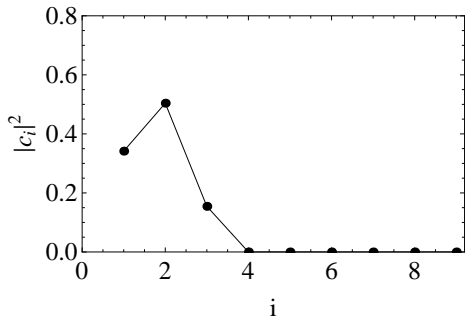}
 \end{center}
 \caption{The optimal values of $\{|c_{i}|^2\}$ for 
 $e^{(2)}=0.06712$ and $n=9$. 
$\{c_i\}$ for $i=4,5,6,7,8,9$ are exactly zero. 
The straight line in  Fig. \ref{fig:two2}(a) is achieved by 
a mixture of this state and the state (with the higher error rate, depicted by the dashed line) shown
 in 
Fig.~\ref{fig:two-amp1}.
\label{fig:two-amp2}}
\end{figure}

In order to clarify the nature of Eve's optimal strategy, 
let us consider a state on Alice's and Bob's quantum registers $AB$,
which is written in the form (see Eq.~(\ref{eq:opt-state1}))
\begin{eqnarray}
 \sum_i c_{i} \ket{\vec{a}+\vec{b}_i}_A\otimes \ket{i}_B.
\label{eq:opt-state2}
\end{eqnarray}
Fig. \ref{fig:two-amp1} shows the amplitudes $\{|c_i|^2\}$
for examples of states achieving the maximal phase error
in the low-error regime with $\vec{a}=0100\ldots 0$.
Mathematically, we see from Eq.~(\ref{eq:koashi-pipha})
that Eve wants to give larger weights on $|c_1|^2$ through $|c_3|^2$
to increase the phase error rate. On the other hand, 
as Eq.~(\ref{eq:condition_e_0}) shows, she must keep 
$\{|c_i|^2\}$ almost uniform in order to avoid the increase of the 
bit errors. As a result, for $e^{(2)}$ almost zero, the distribution 
$\{|c_i|^2\}$ is almost uniform, and it starts to cluster around 
$i=2$ when $e^{(2)}$ increases. Note that the values at
the edges $|c_1|^2$ and $|c_9|^2$ are halved due to the particular 
definition of Bob's quantum register $B$ (see Fig.~\ref{DPS}).
We may give a rough physical interpretation of this behavior
as follows. The amplitude $c_i$ represents the event where
 Alice emitted 
a photon in the second pulse and another in the $i$-th pulse, 
and Bob received a photon in the $i$-th pulse. This means that 
Eve has stolen the photon in the second pulse in a PNS attack.
Obviously, Eve gains nothing if Bob receives a photon 
far from the second pulse ($i\ge 4$), and so Eve should decrease the 
rate of such events as best as possible while keeping the bit error rate low.

Similarly, optimal amplitudes for the high error regime is shown in
Fig. \ref{fig:two-amp2}. The bit error rate is chosen for the point where 
the straight line in Fig.~\ref{fig:two2}(a) meets the curve specified by 
$\Omega_+^{(2)}(\lambda)$. For $\vec{a}=1110\ldots 0$,  
 only $c_1$, $c_2$, and $c_3$ can be nonzero, implying that the 
distribution is far from the uniform one and the bit error rate is high.
As discussed in the case of $\nu=1$, this can be understood as Eve tries
to measure the relative phase between $i=1$ and $i=2$ or 
 between $i=2$ and $i=3$.


\subsection{Three-photon part and comparison}

\begin{figure}
\begin{center}
 \includegraphics[scale=0.6]{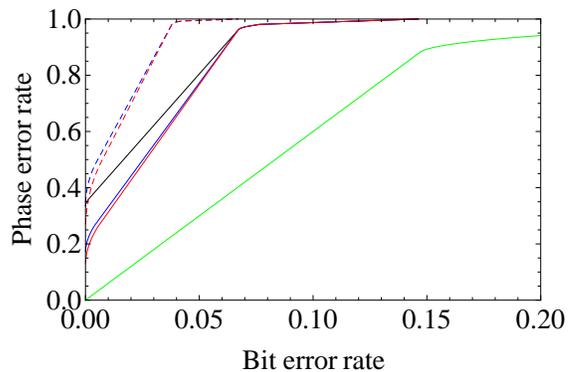}
 \end{center}
 \caption{Color online: Comparison of the upper bounds on the phase
 error rates for one-, two-, and three-photon parts. 
The black, blue, and red solid curves are bounds for $\nu=2$, 
respectively corresponding to
 $n=4$, $n=7$, and $n=9$.
The blue and red dashed curves are bounds for $\nu=3$, 
respectively corresponding to $n=7$, and $n=9$.
(All the region is achievable for $n=4$ and $\nu=3$.)
The green curve represents the bound for $\nu=1$,
which is independent of $n$.
\label{fig:compare}}
\end{figure}

The case for $\nu=3$ can be calculated in a similar manner as for $\nu=2$,
and it was confirmed that the boundary is achieved 
by states with $\vec{a}=11110\ldots 0$ and 
$\vec{a}=0110\ldots 0$, or a mixture of them.
We compare the upperbounds on the phase error rate 
for $\nu=0,1,2,3$ (see Fig.~\ref{fig:compare}).
The bound for $\nu=1$ is independent of $n$. 
No bound is shown for $n=4$ and $\nu=3$ since all the region is 
achievable in this case.
The achievable region for $\nu=0$ is 
a point $(1/2,1/2)$ and is not shown in 
the figure.  
This explicitly verifies $\Omega^{(0)}(\lambda)\le\Omega^{(1)}(\lambda)\le\Omega^{(2)}(\lambda)\le\Omega^{(3)}(\lambda)$ (see Eq. (\ref{chain})).


\section{Key generation rates}\label{keyrates}

\begin{figure}
\begin{center}
 \includegraphics[scale=0.5]{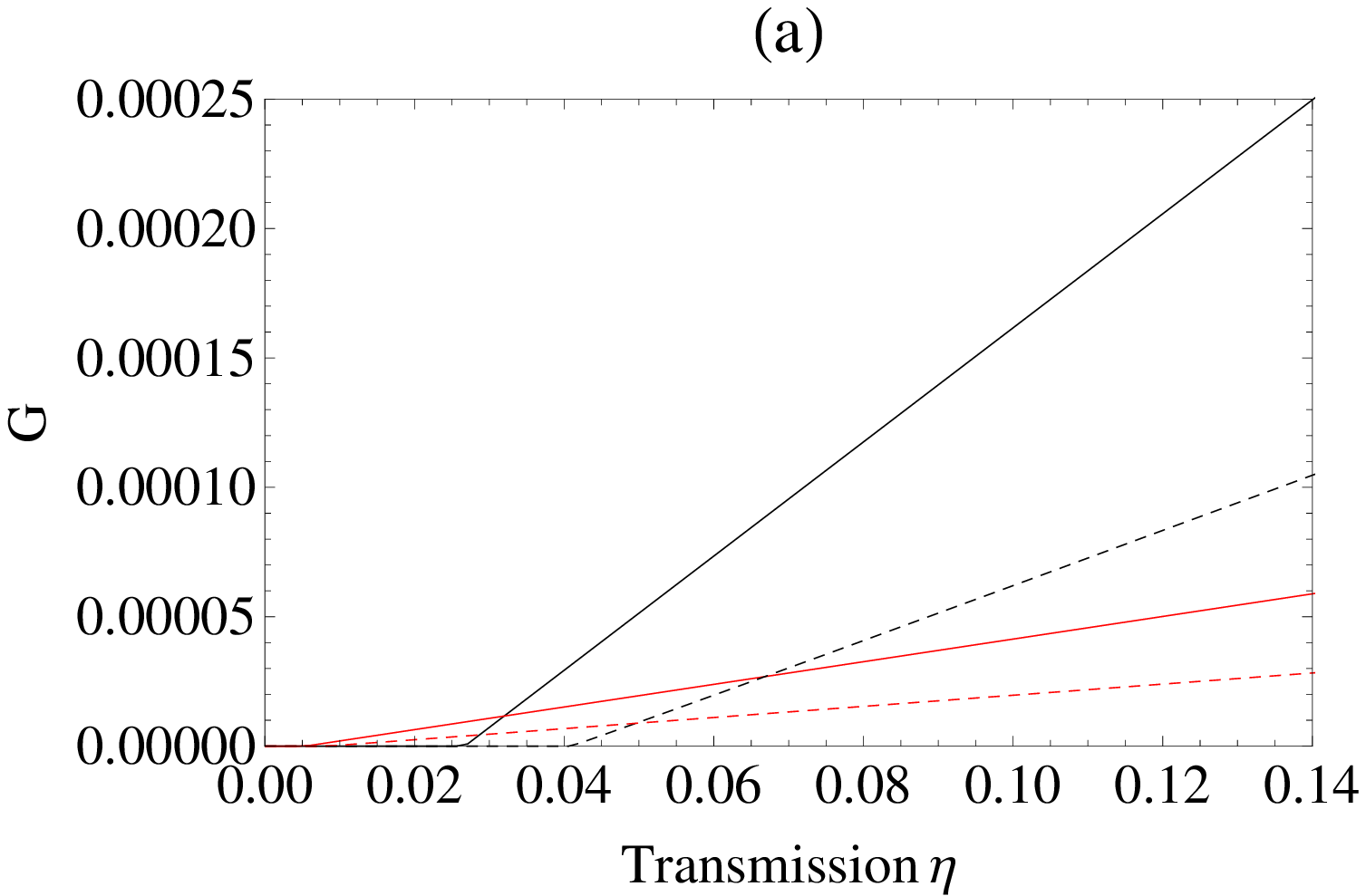}
 \includegraphics[scale=0.5]{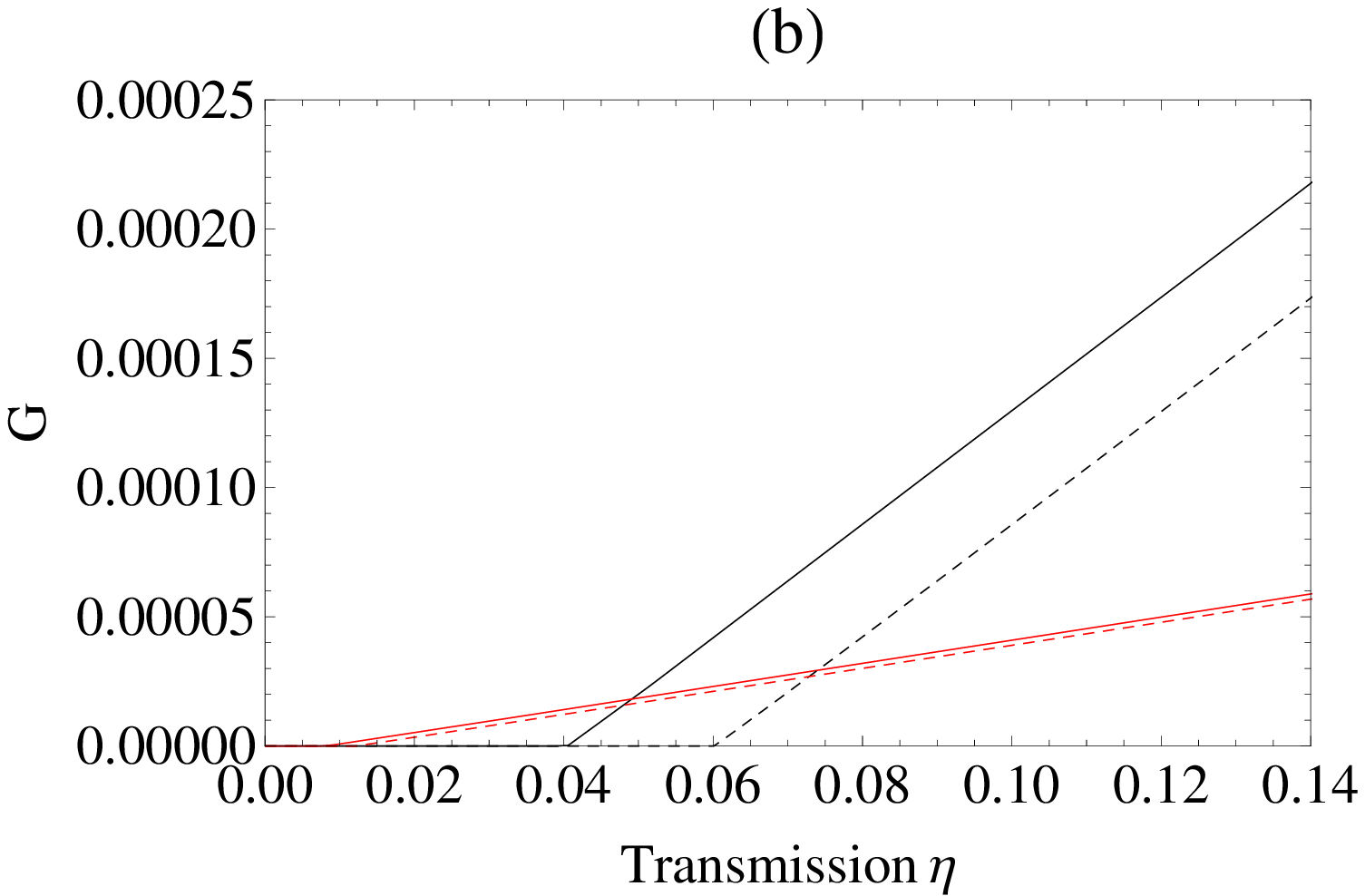}
 \end{center}
 \caption{Color online: The key generation rate as a function of $\eta$
 for (a) $n=4$ and (b) $n=9$. The bit error rate is set to
 $e=3\%$. The solid lines are the rates considering the contribution 
up to the three photon parts (${\overline \nu}=3$),
and the dashed lines use only the single-photon part
(${\overline \nu}=1$).
The mean photon number per block $n\alpha^2$ is either $0.02$ (black)
or $0.004$ (red).
 \label{key1}}
\end{figure}

\begin{figure}
\begin{center}
 \includegraphics[scale=0.55]{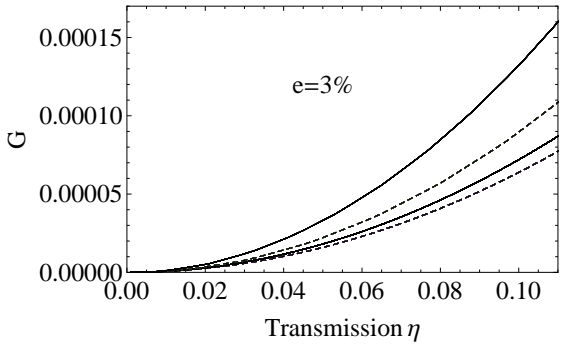}
 \end{center}
 \caption{The key generation rate as a function of $\eta$ for $n=4$ and
 $n=9$, with the mean photon number $n\alpha^2$ optimized for each value 
of $\eta$.  The bit error rate is set to
 $e=3\%$. The solid curves are the rates considering the contribution 
up to the two photon parts (${\overline \nu}=2$),
and the dashed curves use only the single-photon part
(${\overline \nu}=1$). The two lower curves are for $n=4$.
\label{keyopt}}
\end{figure}

In this section, we show examples of the key generation rates by
using a model of the transmission channel. 
We assume an optical channel with linear losses, which are characterized
 by a single-photon transmission rate of $\eta$.
As a source of errors, we assume only alignment errors which lead
to a constant error rate $e$ for every time slot, regardless of 
the value of $\eta$.
With this model, we may use $e$ directly in the theory, and 
the detection rate  $Q$ that we need for calculating the key rate 
is given by
\begin{eqnarray} 
Q&=&(n-1)\eta\alpha^2 e^{-(n+1)\eta\alpha^2}\,,
\label{eq:Qfrometa}
\end{eqnarray}
reflecting the probability of $(n-1)$ time slots receiving one photon in
total and $2$ time slots receiving zero photon.

In Fig. \ref{key1}, we plot the key generation rate for $e=3\%$ and $n=4,9$ as a function of $\eta$. We picked up 
two cases of the mean photon number of the pulses,
$n\alpha^2=0.004$ and $n\alpha^2=0.02$.
In each case, we apply two types of analyses to calculate the key rate.
The first one is to consider the contribution from the single-photon,
the two-photon, and the three-photon part, 
which we denote by ${\overline \nu}=3$.
The other one is to consider contribution from only the single-photon
part, 
which we denote by ${\overline \nu}=1$. 
The detail of each analysis is described in Appendix~\ref{app:calc_key}.
We can see that for each $n$ and $n\alpha^2$, the key generation from ${\overline \nu}=3$ is larger than the one from ${\overline \nu}=1$, which indicates the positive key generation from multiple photons.

In order to assert that contribution from multi-photon parts 
do improve the key rate, we compare the key rates after the 
optimization over the mean photon number at each value of $\eta$.
For this purpose, we limit ourselves to the analysis up to 
the two-photon parts (${\overline \nu}=2$), and compare the rates to
those from the ${\overline \nu}=1$ analysis. The results for 
$n=4, 9$ with $e=3\%$ are shown in Fig. \ref{keyopt}.
We see clearly that the rate from the ${\overline \nu}=2$ analysis
exceeds the best rate from the ${\overline \nu}=1$ analysis,
which confirms a positive contribution from the two-photon part.
The ratio of the key rate for $\overline\nu=2$ to that for 
$\overline\nu=1$ is
almost constant over $\eta$, and is about
$1.1$ for $n=4$ and $1.5$ for $n=9$.   
We have confirmed that 
the optimal mean photon number for each case increases almost linearly
as $\eta$ increases, i.e., $n\alpha^2\cong D(\overline{\nu})\eta$
with a constant $D(\overline{\nu})$ that depends on the types
of analyses. The ratio $D(\overline{\nu}=2)/D(\overline{\nu}=1)$
is about $1.4$ for $n=9$ and $1.1$ for $n=4$.
Hence the inclusion of the two photon part in the analysis 
does not only give an extra portion of the key from the two-photon 
events, but also allows us to use more intense pulses to increase 
the detection rate itself to improve the key rate.

\begin{figure}
\begin{center}
 \includegraphics[scale=0.52]{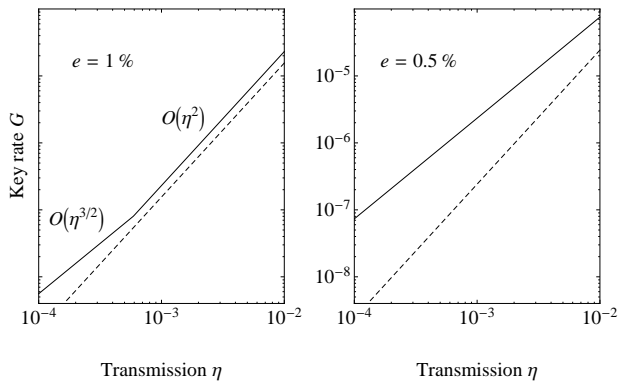}
 \end{center}
 \caption{The key generation rate $G$ for $n=9$
as a function of 
transmission $\eta$ when the error rate $e$ is low, 
for $e=1\%$ (left) and $e=0.5\%$ (right).
The solid curves are the rates considering the contribution 
up to the two photon parts (${\overline \nu}=2$),
and the dashed curves use only the single-photon part
(${\overline \nu}=1$).
For the dashed curves, the mean photon number $n\alpha^2$ 
is optimized to maximize the rate $G$.
For the solid curves, the mean photon number
is chosen as 
either $n\alpha^2\cong 0.0987\eta$ or 
$n\alpha^2\cong 0.0465\sqrt \eta$ for $e=1\%$,
and $n\alpha^2\cong 0.105\sqrt \eta$ for $e=0.5\%$.
\label{logplot}}
\end{figure}

When the error rate is lower than $\sim 1\%$, the difference between 
the rate from the multi-photon analysis 
and that from the single-photon analysis 
becomes more pronounced, as shown in the logarithmic plots
in Fig.~\ref{logplot} for $n=9$.
For $e=1\%$, the key rate 
shows an $O(\eta^2)$ dependence when $\eta$ is 
higher, and the case of ${\overline \nu}=2$ is better than the case of
${\overline \nu}=1$ by a constant factor. For this rate, 
the mean photon number is chosen as 
$n\alpha^2=O(\eta)$.
This behavior is the same as that for
$e=3\%$. But when $\eta$ is lower, the rate starts to show
 an improved dependence of $O(\eta^{3/2})$. This is due to 
the fact that it is possible to generate the final key solely from
the two-photon events when $e=1\%$, and hence 
we have another option of choosing the mean photon number $n\alpha^2$
to be $O(\sqrt{\eta})$, much higher than in the case of $e=3\%$.
When $e=0.5\%$, the $O(\eta^{3/2})$ dependence already shows up 
at $\eta$ as high as $0.01$, and the multi-photon rate is 
much higher than the single-photon rate.
This types of behavior can be also analytically explained in the limit of 
$\eta\to 0$, which is given in Appendix~\ref{appendix-low-limit}.

\section{Summary and discussion}\label{summary}

In this paper, we have proved the unconditional security of
coherent-state-based DPS protocol with block-wise phase
randomization. Thanks to the block-wise phase randomization, Alice's
state is transformed to a classical mixture of
states with fixed numbers of the total photons, 
which allows us to analyze each photon-number space
separately.
After the calculation of the bounds on the 
phase error rate for the single-photon part, the
two-photon part, and the three-photon part,
we showed behavior of the key generation rates 
assuming a simple channel model that accommodates
losses and misalignment. 

The examples of the key generation rates first show that 
DPS with a weak coherent pulse train 
is able to produce a secret key with no assumptions on the 
ability of the eavesdropper. 
When
the bit error rate is relatively large (about $3\%$), 
the key generation rate is proportional to $O(\eta^{2})$ 
of the channel transmission $\eta$. The multi-photon emission 
events contributes to the key rate as an improvement 
of a constant factor.
When the bit error rate is small (about $1\%$ or smaller), 
an improved  scaling, $O(\eta^{3/2})$, of the key rate 
is observed, which implies the key can be generated solely from 
the two-photon emission 
events. These behaviors confirm the expected 
robustness of the DPSQKD protocol against the PNS attacks. 

We remark that our analysis does not fully exploit the available
data. For instance, we have used the bit error rate averaged over
 the block for the security analysis, but the bit error rate for
each time slot is also available in the experiments. Similarly, 
our analysis
does not take into account the detection rate for each time
slot. Moreover, for the simplicity of analysis, we considered one-photon
detection per block by Bob, and it will be interesting to consider multiple
detection of a single photons in a block. We believe that the security
proof that fully makes use of all of them improves the key generation
rate, which we leave for future study. We hope that the ideas introduced this
paper is useful for the security proof of the so-called COW protocol,
which shares some similarities with DPSQKD.

\section*{Acknowledgment}
We thank Daniel Gottesman, Norbert L{\"u}tkenhaus, Hoi-Kwong Lo, Marcos Curty, Toshimori Honjo, Hiroki Takesue, Yasuhiro Tokura, Yodai Watanabe, and Koji Azuma for valuable comments and enlightening discussions. This research is in part supported by the project ``Secure photonic network technology'' as part of ``The project UQCC'' by the National Institute of Information and Communications Technology (NICT) of Japan and in part by the Japan Society for the Promotion of Science (JSPS) through its Funding Program for World-Leading Innovative R$\&$D on Science and Technology (FIRST Program)".

\appendix
\section{Calculation of the key rate}
\label{app:calc_key}

Here we describe the procedure we used to determine the key rates shown
in Sec.~\ref{keyrates}. 
Instead of calculating the bound Eq.~(\ref{eq:koashi-6}) exactly, 
we used weaker bounds in which we neglect the contribution of 
the events where the photon number $\nu$ in 
Alice's pulse train is larger than a constant $\overline{\nu}$,
namely, the contribution is considered up to $\overline{\nu}$ photons.
This was done by assuming 
\begin{eqnarray}
 \Omega^{(\nu)}(\lambda)=1 \;\; (\nu \ge \overline{\nu}+1),
\end{eqnarray}
which means that 
\begin{eqnarray}
 \Omega^{(\nu)}_{h}(\gamma)=1 \;\; (\nu \ge \overline{\nu}+1).
\end{eqnarray}
These choices trivially satisfy
Eqs.~(\ref{eq:koashi-2}) and (\ref{eq:koashi-4})
for any $\lambda\ge 0$ and $\gamma\ge 0$.

Three types of analyses, $\overline{\nu}=1,2$ and $3$, were used 
in Sec.~\ref{keyrates}. 
The condition of Eq.~(\ref{chain}) is now written as
\begin{eqnarray}
 \Omega^{(0)}(\lambda) \le \Omega^{(1)}(\lambda) \le \cdots
\le \Omega^{(\overline{\nu})}(\lambda)
\le 1,\nonumber\\
\end{eqnarray}
which was confirmed to be true in Sec. VB for $\overline{\nu}=1,2,3$.
This assures that, under the given values of $Q$ and $e$,
 the maximum value of 
\begin{eqnarray}
 h^{({\rm ph})}=\sum_{\nu=0}^\infty q^{(\nu)} h(e^{({\rm ph}, \nu)})
\end{eqnarray}
is the same as the maximum value of 
\begin{eqnarray}
 \tilde{h}^{({\rm ph})}\equiv
\sum_{\nu=0}^\infty q^{(\nu)*} h(e^{({\rm ph}, \nu)}),
\end{eqnarray}
where $\{q^{(\nu)*}\}$ are
defined in Eq.~(\ref{qu}). Let us define 
\begin{eqnarray}
 q^{(\nu>\overline{\nu})*}\equiv \sum_{\nu=\overline{\nu}+1}^\infty q^{(\nu)*}, 
\end{eqnarray}
and assume $q^{(0)*}=0$, which is satisfied by normally observed values
of $Q$. Then we have  
\begin{eqnarray}
 \tilde{h}^{({\rm ph})}\le q^{(\nu>\overline{\nu})*}+
\sum_{\nu=1}^{\overline{\nu}} q^{(\nu)*} h(e^{({\rm ph}, \nu)}).
\end{eqnarray}
We will then explain the evaluation of the right-hand side
for each value of $\overline{\nu}$.

For $\overline{\nu}=1$, we use $e\ge q^{(1)*}e^{(1)}$
and Eq.~(\ref{eq:6e}) to obtain
\begin{eqnarray}
 h(e^{({\rm ph}, 1)})\le h(6 e/ q^{(1)*})
\end{eqnarray}
for $12 e\le q^{(1)*}$. Hence we obtain 
\begin{eqnarray}
 h^{({\rm ph})}\le q^{(\nu>1)*}+ q^{(1)*}h(6 e/ q^{(1)*})
\;\; (12 e\le q^{(1)*}).
\end{eqnarray}

For $\overline{\nu}=2$, we start from 
\begin{eqnarray}
 e^{({\rm ph}, 2)} \le \Omega^{(2)}(\lambda)+\lambda e^{(2)},
\end{eqnarray}
we see that for any $\lambda\ge 0$ and
$\tilde{e}$ such that 
$0 \le \Omega^{(2)}(\lambda)+\lambda \tilde{e} \le 1/2$,
we have
\begin{eqnarray}
h( e^{({\rm ph}, 2)})\le 
h(\Omega^{(2)}(\lambda)+\lambda \tilde{e})
+ \gamma^*(e^{(2)} - \tilde{e})
\end{eqnarray}
with 
\begin{eqnarray}
 \gamma^*\equiv h'(\Omega^{(2)}(\lambda)+\lambda \tilde{e})
\ge 0.
\end{eqnarray}
Combining it with Eq.~(\ref{eq:koashi-4}), we obtain 
\begin{eqnarray}
q^{(1)*} h(e^{({\rm ph}, 1)})
+
q^{(2)*} h(e^{({\rm ph}, 2)})
\le f(\lambda, \tilde{e})
\end{eqnarray}
with
\begin{multline}
f(\lambda, \tilde{e})\equiv 
\gamma^* e+q^{(1)*}\Omega_h^{(1)}(\gamma^*)
\\
 +q^{(2)*}\left[
h(\Omega^{(2)}(\lambda)+\lambda \tilde{e})
- \gamma^* \tilde{e})
\right],
\end{multline}
which leads to a bound
\begin{eqnarray}
 h^{({\rm ph})} \le f(\lambda, \tilde{e})+ q^{(\nu>2)*}.
\end{eqnarray}
The evaluation of $f(\lambda, \tilde{e})$ is not difficult 
since it is easy to calculate 
$\Omega_h^{(1)}(\gamma^*)$ for a given numerical value of $\gamma^*$
through the use of Eq.~(\ref{eq:koashi-omegah1gamma}).
In order to obtain the best bound, it is sufficient to try the following 
combinations,
\begin{eqnarray}
 \{(\lambda, \tilde{e})|\lambda\ge 0,
 \tilde{e}_{+}(\lambda)
\le \tilde{e} \le \tilde{e}_{-}(\lambda)
\}
\end{eqnarray}
with
\begin{eqnarray}
 \tilde{e}_\pm(\lambda)\equiv 
- \lim_{\lambda_0\to \lambda\pm 0}
\frac{d\Omega^{(2)}(\lambda_0)}{d\lambda_0},
\end{eqnarray}
which is essentially a one-parameter family.

For $\overline{\nu}=3$, we define 
\begin{eqnarray}
&& 
 q^{(2,3)*}\equiv q^{(2)*}+ q^{(3)*},
\nonumber \\
&& 
r_2\equiv q^{(2)*}/q^{(2,3)*}, 
\;\;
r_3\equiv q^{(3)*}/q^{(2,3)*},
\nonumber \\
&& 
e^{({\rm ph}, 2,3)}\equiv  r_2 e^{({\rm ph}, 2)} 
+r_3 e^{({\rm ph}, 3)},
\nonumber \\
&& 
e^{(2,3)}\equiv  r_2 e^{(2)} 
+r_3 e^{(3)},
\nonumber \\
&& 
\Omega^{(2,3)}(\lambda)\equiv 
r_2 \Omega^{(2)}(\lambda)
+r_3 \Omega^{(2)}(\lambda),
\end{eqnarray}
and introduce another compromise in the key rate 
by the use of the inequality
\begin{multline}
 q^{(2)*}h(e^{({\rm ph}, 2)})+q^{(3)*}h(e^{({\rm ph}, 3)})
\le q^{(2,3)*} h( e^{({\rm ph}, 2,3)}),
\end{multline}
which is not necessarily saturated.
Starting from 
\begin{eqnarray}
 e^{({\rm ph}, 2,3)} \le \Omega^{(2,3)}(\lambda)+\lambda e^{(2,3)},
\end{eqnarray}
we can exactly follow the argument for $\overline{\nu}=2$,
except that $q^{(2)*}$, $\Omega^{(2)}$, and $q^{(\nu>2)*}$
  should be replaced with $q^{(2,3)*}$, $\Omega^{(2,3)}$, and $q^{(\nu>3)*}$,
respectively.

\section{Limit of low channel transmission}\label{appendix-low-limit}

Here we discuss the behavior of the key rate in the limit of 
low channel transmission, $\eta\to 0$.
For comparison, first we consider the key rate under the 
assumption that only  
the single-photon events ($\nu=1$) contribute the final key, namely, 
\begin{eqnarray}
 \Omega^{(\nu)}(\lambda)=1 \; \text{for} \; \nu\ge 2.
\end{eqnarray}
In this case, the amplitude must be at most proportional to $\eta$
to mitigate the PNS attacks, resulting in a rate of raw key proportional 
to $\eta^2$. To investigate this quadratic behavior of the key, 
we introduce parameters as follows,
\begin{eqnarray}
 \alpha^2 &=& C_2 \eta
\\
 G &=& D_2 \eta^2,
\end{eqnarray}
where the subscript $2$ implies the quadratic behavior.
Taking only the dominant terms in the limit of $\eta\to 0$
in Eqs.~(\ref{eq:koashi-3}), (\ref{qu}),~(\ref{eq:koashi-6}), (\ref{eq:Qfrometa}), we have
\begin{eqnarray}
 Q&=& (n-1) C_2 \eta^2 
\\
 y &\equiv& q^{(1)*}= 1-\frac{(n\alpha^2)^2}{2Q}=1-\frac{n^2 C_2}{2(n-1)} \label{C2y}
\\
 D_2&=& (n-1) C_2 \left\{
y[1-\Omega^{(1)}_h (\gamma)]-\gamma e-h(e) 
\right\},
\end{eqnarray}
where we restrict the range of $C_2$ to be
\begin{eqnarray}
 0\le C_2 \le \frac{2(n-1)}{n^2}.
\label{eq:C2range}
\end{eqnarray}
Setting $\gamma=6h'(6e/y)$ and using  Eqs.~(\ref{C2y}) and (\ref{eq:koashi-omegah1gamma}),
we obtain
\begin{eqnarray}
 D_2= \frac{2(n-1)^2}{n^2}(1-y)\left[
y-yh\left(\frac{6e}{y}\right)-h(e)
\right]
\end{eqnarray}
Hence $D_2$ can be as large as the following quantity
\begin{multline}
 D_2^{(1)}(n,e)\equiv \\
\frac{2(n-1)^2}{n^2}\max_{0\le y \le 1}
(1-y)\left[
y-yh\left(\frac{6e}{y}\right)-h(e)
\right],
\end{multline}
where the superscript $1$ signifies the contribution from only the
single-photon events.
Let us define a constant $e_{\rm max}^{(1)}\cong 0.0375$ by the equation
\begin{eqnarray}
  1-h(6e_{\rm max}^{(1)})-h(e_{\rm max}^{(1)})=0.
\end{eqnarray}
For $e<e_{\rm max}^{(1)}$, $D_2^{(1)}(n,e)>0$ and we have a positive key
rate $G=D_2^{(1)}(n,e) \eta^2$.

\begin{figure}
\begin{center}
 \includegraphics[scale=0.8]{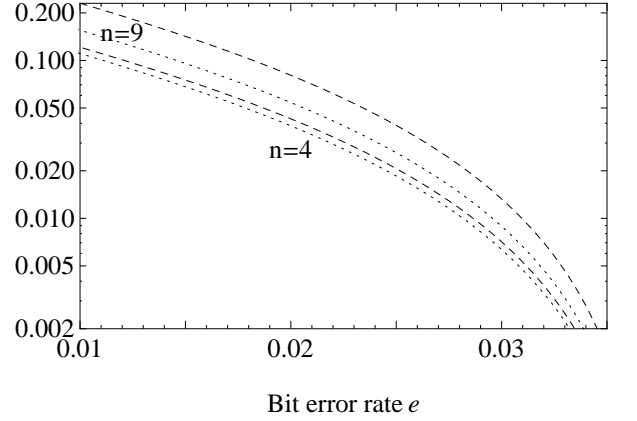}
 \end{center}
 \caption{Comparison of the key rates in the limit of low transmission
 $\eta\to 0$, when the contribution from two-photon emission events 
are taken into account (the dashed curves) or not (the dotted curves).
 When the total mean photon number $n\alpha^2$ of
 the $n$ pulses is chosen to scale as $O(\eta)$, a key rate of 
$G=O(\eta^2)$ is achievable. If we assume that the key is 
contributed only from  
the events where Alice has emitted a single photon,
the key rate scales as 
$G\sim D^{(1)}_{2}(n,e) \eta^2$, with the coefficient 
$D^{(1)}_{2}(n,e)$ being plotted as dotted curves for $n=4$ and $9$.
When we include the contribution from the events where 
Alice has emitted two photons, the key rate is improved to 
$G\sim D^{(1,2)}_{2}(n,e) \eta^2$, and 
the coefficient $D^{(1,2)}_{2}(n,e)$ 
is shown as the dashed curves.
\label{fig:apx2}}
\end{figure}

Next, let us include the contribution from the two-photon events
$(\nu=2)$ by using $\Omega^{(2)}(\lambda)$ calculated in
Sec.~\ref{sec:two-photon}
and $\Omega^{(2)}_h(\gamma)$ that can be calculated from
$\Omega^{(2)}(\lambda)$.
In this case, we have
\begin{multline}
  D_2= \frac{2(n-1)^2}{n^2}(1-y) \left\{y[1-\Omega^{(1)}_h (\gamma)] \right.
\\
\left. 
+(1-y)[1-\Omega^{(2)}_h (\gamma)]-\gamma e-h(e) 
\right\}.
\end{multline}
As a function of $y$, $D_2$ is maximal at $y=y^*(\gamma)$ with
\begin{eqnarray}
 y^*(\gamma)\equiv 1-
\frac{1-h(e)-\gamma e - \Omega^{(1)}_h(\gamma)}
{2[\Omega^{(2)}_h(\gamma)-\Omega^{(1)}_h(\gamma)]}
\end{eqnarray} 
when $0\le y^*(\gamma) \le 1$.
For $n\ge 11$, let $e=e_{\rm min}^{(1,2)}$ be the solution of the following set of 
equations,
\begin{eqnarray}
 \frac{1-h(e)-\gamma e - \Omega^{(1)}_h(\gamma)}
{2[\Omega^{(2)}_h(\gamma)-\Omega^{(1)}_h(\gamma)]}&=& 1
\\
 \frac{d\Omega^{(2)}_h(\gamma)}{d\gamma}&=&-e,
\end{eqnarray}
which depends on $n$. For $n\le 10$, there is no positive solution 
and let $e_{\rm min}^{(1,2)}\equiv 0$.
Then it can be shown that 
for $e_{\rm min}^{(1,2)}\le e<e_{\rm max}^{(1)}$, we have a positive key 
rate $G=D_2^{(1,2)}(n,e) \eta^2$ with 
\begin{multline}
  D_2^{(1,2)}(n,e)\equiv \frac{(n-1)^2}{n^2} \times
\\
\max_{\gamma\ge 0:0\le y^*(\gamma)\le 1}
[1-y^*(\gamma)]
[1-h(e)-\gamma e - \Omega^{(1)}_h(\gamma)]
\end{multline}
where it should be noted that $\Omega^{(2)}_h(\gamma)$ 
depends also on $n$.
For $e<e_{\rm min}^{(1,2)}$, there is no maximal point of $C_2$
in the range of Eq.~(\ref{eq:C2range}) and the key rate 
increases indefinitely with $C_2$, implying that the scaling 
is better than $O(\eta^2)$, which will be confirmed below.
The comparison between $D_2^{(1)}(n,e)$ and $D_2^{(1,2)}(n,e)$ for 
$n=4,9$ is shown in Fig.~\ref{fig:apx2}. 
The ratio $D_2^{(1,2)}/D_2^{(1)}$ is only weakly dependent on 
the bit error rate $e$ and approximately 1.1 and 1.5 for
$n=4$ and $n=9$, respectively.

\begin{figure}
\begin{center}
 \includegraphics[scale=0.8]{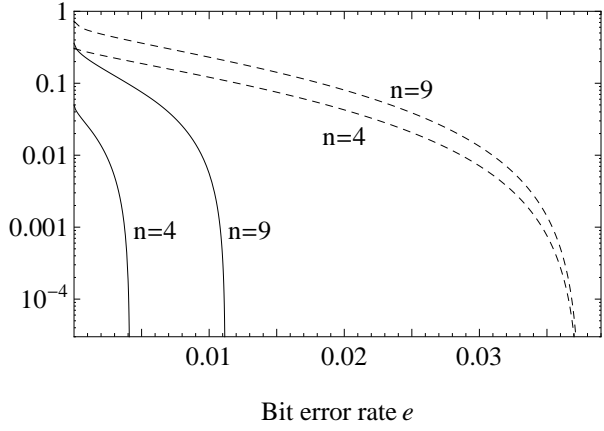}
 \end{center}
 \caption{Coefficients of the key rate $G$ in the limit of low transmission
 $\eta\to 0$. By choosing the total mean photon number $n\alpha^2$ of
 the $n$ pulses to scale as $O(\eta)$, a key rate of 
$G\sim D^{(1,2)}_{2}(n,e) \eta^2$ is achievable. The coefficient
 $D^{(1,2)}_{2}(n,e)$ is shown for $n=4,9$ as the broken curves.
 When $e$ is small, it is also possible to choose 
$n\alpha^2=O(\sqrt{\eta})$, resulting in a better scaling of the key
 rate as $G\sim D^{(2)}_{3/2}(n,e) \eta^{3/2}$. 
The coefficient $D^{(2)}_{3/2}(n,e)$ is shown for $n=4,9$ as the solid curves. 
\label{fig:apx1}}
\end{figure} 

When the bit error rate is as small as $\sim 1\%$, 
it is also possible to extract the key solely from the two-photon 
part. This means that $\alpha^2$ is no longer required to be 
proportional to $\eta$. Instead, we may obtain a positive key rate
with a higher amplitude as
\begin{eqnarray}
 \alpha^2 &=& C_{3/2} \eta^{1/2}
\\
 G &=& D_{3/2} \eta^{3/2}, 
\end{eqnarray} 
with an improved dependence on the channel transmission $\eta$.
For simplicity, we still ignore the contribution from the events 
with $\nu\ge 3$, namely, we assume
\begin{eqnarray}
 \Omega^{(\nu)}(\lambda)=1 \; \text{for} \; \nu\ge 3.
\end{eqnarray}
Assuming $C_{3/2}^2\le 6(n-1)/n^3$ and 
taking dominant terms in the limit of $\eta\to 0$, we have $q^{(1)*}=0$ and 
\begin{eqnarray}
 Q&=& (n-1) C_{3/2} \eta^{3/2} 
\\
 z &\equiv& q^{(2)*}= 1-\frac{(n\alpha^2)^3}{6Q}=1-\frac{n^3 C_{3/2}^2}{6(n-1)}
\\
 D_{3/2}&=& \sqrt{6}\left(\frac{n-1}{n}
\right)^{3/2} (1-z)^{1/2}\times
\nonumber \\
&& \left\{
z[1-\Omega^{(2)}_h (\gamma)]-\gamma e-h(e) 
\right\}.
\end{eqnarray} 
As a function of $z$, $D_{3/2}$ is maximal at $z=z^*(\gamma)$ with
\begin{eqnarray}
 z^*(\gamma)\equiv 1-
\frac{1-h(e)-\gamma e - \Omega^{(2)}_h(\gamma)}
{3[1-\Omega^{(2)}_h(\gamma)]}
\end{eqnarray} 
when $0\le z^*(\gamma) \le 1$. 
Let $e=e_{\rm max}^{(2)}(n)$ be the solution of the following set of 
equations,
\begin{eqnarray}
 1-h(e)-\gamma e - \Omega^{(2)}_h(\gamma)&=&0
\\
 \frac{d\Omega^{(2)}_h(\gamma)}{d\gamma}&=&-e,
\end{eqnarray}
which depends on $n$. 
Examples of the values are $e_{\rm max}^{(2)}(4)\cong 0.0041$
and $e_{\rm max}^{(2)}(9)\cong 0.0112$.
Then,
for $e<e_{\rm max}^{(2)}(n)$, we have a positive key 
rate $G=D_{3/2}^{(2)}(n,e) \eta^{3/2}$ with 
\begin{multline}
  D_{3/2}^{(2)}(n,e)\equiv \frac{2\sqrt{6}}{3}\left(\frac{n-1}{n}
\right)^{3/2} \times
\\
\max_{\gamma\ge 0:0\le y^*(\gamma)\le 1}
[1-z^*(\gamma)]^{1/2}
[1-h(e)-\gamma e - \Omega^{(2)}_h(\gamma)].
\end{multline}
In Fig.~\ref{fig:apx1}, $D_{3/2}^{(2)}(n,e)$ and $D_{2}^{(1,2)}(n,e)$
are shown for $n=4$ and $n=9$.


\begin{thebibliography}{}

\bibitem{maybers96} D. Mayers, Lect. Notes Comput. Sci. {\bf 1109}, 343 (1996).

\bibitem{LC} H.-K. Lo and H. F. Chau, Science {\bf 283}, 2050 (1999). 

\bibitem{SP} P. W. Shor and J. Preskill, Phys. Rev. Lett. {\bf 85}, 441 (2000).

\bibitem{GLLP} D. Gottesman, H.-K. Lo, N. L$\ddot{\mbox{u}}$tkenhaus, and J. Preskill, Quantum Information and Computation {\bf 5}, 325 (2004).

\bibitem{EDP} C. H. Bennett, D. P. DiVincenzo, J. A. Smolin, and W. K. Wootters, Phys. Rev. A {\bf 54}, 3824 (1996).

\bibitem{TKI04} K. Tamaki, M. Koashi, and N. Imoto, Phys. Rev. Lett. {\bf 90}, 167904 (2003).



\bibitem{Azuma} K. Azuma, T$\overline{\rm o}$hoku Math. J. {\bf 19} 357 (1967).

\bibitem{Azuma2} J.-C. Boileau, K. Tamaki, J. Batuwantudawe, R. Laflamme, and, J. M. Renes, Phys. Rev. Lett. {\bf 94} 040503 (2005), K. Tamaki and H.-K. Lo, Phys. Rev. A. {\bf 73}, 010302(R) (2006), K. Tamaki, N. L$\ddot{\mbox{u}}$tkenhaus, M. Koashi, J. Batuwantudawe, Phys. Rev. A {\bf 80} 032302 (2009). 


\bibitem{Renner} C. M. Caves, C. A. Fuchs, J. Math. Phys. {\bf 43}, 4537 (2002). R. Koenig, and Renato Renner, J. Math. Phys. {\bf 46}, 122108 (2005).

\bibitem{DLM06} See for instance, M. Dusek, N. L\"utkenhaus, and M. Hendrych, Progress in Optics, Vol. {\bf 39}, 381, Edt. E. Wolf, Elsevier (2006), ArXiv: quant-ph/0601207.

\bibitem{DPS} K. Inoue, E. Waks, and Y. Yamamoto, Phys. Rev. Lett. \textbf{89}, 037902 (2002), K. Inoue, E. Waks and Y. Yamamoto, Phys. Rev. A \textbf{68}, 022317 (2003).

\bibitem{COW} N. Gisin, G. Ribordy, H. Zbinden, D. Stucki, N. Brunner, V. Scarani, quant-ph/0411022 (2004). D. Stucki, N. Brunner, N. Gisin, V. Scarani, H. Zbinden, Appl. Phys. Lett. 87, 194108 (2005).

\bibitem{COW2} We note that the security of COW protocol with coherent state was proven by the recent paper (T. Moroder, M. Curty, C. C. W. Lim, L. P. Thinh, H. Zbinden, and N. Gisin, ArXiv:1207.5544), and according to the paper, their proof technique is applicable to the security proof of DPSQKD.

\bibitem{single-DPS} W. Kai, K. Tamaki, and Y. Yamamoto, Phys. Rev. Lett. {\bf 103}, 170503 (2009).

\bibitem{composable} R. Renner, and R. Koenig,  Proc. of TCC 2005, LNCS, Springer, vol. 3378 (2005), M. Ben-Or, Michal Horodecki, D. W. Leung, D. Mayers, J. Oppenheim, Theory of Cryptography: Second Theory of Cryptography Conference, TCC (2005). J.Kilian (ed.) Springer Verlag 2005, vol. 3378 of Lecture Notes in Computer Science, 386.
 
\end{thebibliography}
\end{document}